\documentclass[sigplan,screen]{acmart}

\usepackage{silence}
\usepackage{tikz}
\usetikzlibrary{shapes.geometric,arrows.meta,positioning,arrows}

\usepackage{xspace}
\definecolor{keyword}{HTML}{2F7AFF}
\definecolor{identifier}{HTML}{000000}
\definecolor{commentstyle}{HTML}{5E5E5E}
\definecolor{cell1}{HTML}{FF5B00}
\definecolor{cell2}{HTML}{ffa200}
\definecolor{cell3}{HTML}{f9da24}
\definecolor{cell4}{HTML}{f9da24}
\definecolor{cell5}{HTML}{cff09e}
\usepackage{colortbl}

\usepackage{anyfontsize}
\usepackage{pifont}

\usepackage{soul}
\usepackage{lscape}
\usepackage{enumitem}
\usepackage{lipsum}
\usepackage{diagbox}
\usepackage{balance}

\usepackage{subfig}
\usepackage{booktabs}
\usepackage{longtable}
\usepackage{ctable}
\usepackage{multirow}
\usepackage{multicol}
\usepackage{makecell}
\usepackage{array}
\newcolumntype{P}[1]{>{\centering\arraybackslash}p{#1}}
\newcolumntype{R}[1]{>{\raggedleft\arraybackslash}p{#1}}
\newcolumntype{C}[1]{>{\centering\arraybackslash}p{#1}}
\newcolumntype{L}[1]{>{\raggedright\arraybackslash}p{#1}}
\usepackage[ruled,vlined,linesnumbered]{algorithm2e}

\usepackage{listings}
\lstdefinelanguage{mylanguage}{
  keywords={True, False, return, switch, if, in, while, do, else, case, break, def, for},
  keywordstyle=\color{keyword}\bfseries,
  ndkeywords={class, export, boolean, throw, implements, import, this},
  ndkeywordstyle=\color{darkgray}\bfseries,
  identifierstyle=\color{identifier},
  sensitive=false,
  comment=[l]{\#},
  morecomment=[s]{/*}{*/},
  commentstyle=\color{commentstyle}\ttfamily,
  stringstyle=\color{red}\ttfamily,
  morestring=[b]',
  morestring=[b]"
}
\lstdefinestyle{mystyle}{
    language=mylanguage,
    backgroundcolor=\color{white},
    numberstyle=\tiny\color{black},
    basicstyle=\ttfamily\footnotesize,
    breakatwhitespace=false,
    breaklines=true,
    captionpos=b,
    keepspaces=true,
    numbers=left,
    numbersep=3pt,
    stepnumber=1,
    showspaces=false,
    showstringspaces=false,
    showtabs=false,
    tabsize=2,
    morekeywords={entry},
    frame=tb,
}
\usepackage{algcompatible}
\usepackage{MnSymbol}
\usepackage{ragged2e}

\usepackage{amsfonts}
\usepackage{mathrsfs}

\newcommand{\compactparagraph}[1]{\noindent{\textbf{\textit{#1}}}}

\newcommand{\optype}[1]{\tt}

\long\def\hidetext#1{}

\newcommand\nott[1]{\bgroup\let\tt\relax#1\egroup}
\title{\nott{\projectname}: Efficient Low-Bit, Conversion-Hiding LLM Decoding on Mobile GPUs}
\newcommand{\projectname}{{\tt BitLLM}\xspace}

\author{Hangyu Zheng}
\email{hyzheng@uga.edu}
\affiliation{%
  \institution{University of Georgia}
  \city{Athens}
  \state{GA}
  \country{USA}
}

\author{Zhihao Shu}
\email{Zhihao.Shu@uga.edu}
\affiliation{%
  \institution{University of Georgia}
  \city{Athens}
  \state{GA}
  \country{USA}
}

\author{Kunxiong Zhu}
\email{Kunxiong.Zhu@uga.edu}
\affiliation{%
  \institution{University of Georgia}
  \city{Athens}
  \state{GA}
  \country{USA}
}

\author{Minghai Qin}
\email{qinminghai@gmail.com}
\affiliation{
  \institution{Western Digital Research}
  \city{San Jose}
  \state{CA}
  \country{USA}
}

\author{Xiaolong Ma}
\email{xiaolongma@arizona.edu}
\affiliation{%
  \institution{University of Arizona}
  \city{Tucson}
  \state{AZ}
  \country{USA}
}

\author{Gagan Agrawal}
\email{gagrawal@uga.edu}
\affiliation{%
  \institution{University of Georgia}
  \city{Athens}
  \state{GA}
  \country{USA}
}

\author{Wei Niu}
\email{wniu@uga.edu}
\affiliation{%
  \institution{University of Georgia}
  \city{Athens}
  \state{GA}
  \country{USA}
}

\renewcommand{\shortauthors}{Hangyu Zheng et al.}

\copyrightyear{2027}
\acmYear{2027}
\setcopyright{cc}
\setcctype{by}
\acmConference[ASPLOS '27]
{Proceedings of the 32nd ACM International Conference on Architectural Support for Programming Languages and Operating Systems, Volume 1}
{April 11--15, 2027}
{Heraklion, Greece.}
\acmBooktitle{Proceedings of the 32nd ACM International Conference on Architectural Support for Programming Languages and Operating Systems, Volume 1 (ASPLOS '27), April 11--15, 2027, Heraklion, Greece}
\acmISBN{979-8-4007-3012-2/2027/04}
\acmDOI{10.1145/3845814.3850028}
\begin{document}

\begin{abstract}

The rapid scaling of large language models (LLMs) has unlocked remarkable capabilities across various domains.
However, deploying LLMs on mobile devices remains challenging
because of their high computational and memory requirements.
Low-bit quantization (e.g., 4-bit weights, 8-bit activations) can reduce model size while preserving accuracy,
but its practical efficacy on mobile GPUs has been limited so far.
In this paper, we identify that the performance gap stems primarily from
{\em frequent data type conversions} between quantized and floating-point values,
and {\em dynamic and inefficient memory access patterns} caused by
the growing Key-Value (KV) cache during autoregressive decoding.
We introduce {\tt BitLLM}, a mobile inference engine
that addresses type conversion overhead and suboptimal memory accesses in quantized LLMs.
Our approach features an adaptive layout and packing strategy for the KV cache that
ensures efficient memory access throughout LLM decoding.
This is coupled with an integer-native execution pipeline that defers and overlaps type conversion outside the compute {\em hot path}, keeping the critical {\tt MatMul} stages in the integer domain.
Implemented on off-the-shelf mobile GPUs,
{\tt BitLLM} outperforms leading engines, i.e., LLAMA.CPP, MNN-LLM, LiteRT-LM, and MLC-LLM,
achieving average speedups of 
1.78$\times$, 1.73$\times$, 1.58$\times$, and 1.47$\times$, respectively,
across various models and quantization levels.
Under long-context generation, 
{\tt BitLLM} reduces average energy consumption (per token) by up to 88.9\%.

\end{abstract}


\maketitle
\section{Introduction}\label{sec:introduction}

Large language models (LLMs) are now central to modern AI, 
having deeply impacted areas such as natural language processing, multimodal reasoning, 
and human-computer interaction~\cite{vu2024gptvoicetasker, taneja2024jill, guan2023intelligent, li2026graph, li2026same}.
This has driven growing demand to run LLMs on mobile
devices~\cite{saeed2024predictive,miao2022real,shureal}, where on-device inference offers privacy,
lower latency, resilience to connectivity issues~\cite{li2024flexnn,niu2021dnnfusion},
and lower data-center energy costs by shifting inference away from centralized infrastructure. 

However, LLMs are inherently resource-intensive, 
requiring billions of parameters, multi-gigaflop computations per token, 
and substantial memory bandwidth for expanding Key-Value (KV) caches~\cite{xiao2023smoothquant}.
These demands strain the limited computational capacity and memory budget of mobile processors~\cite{shen2024agile,xu2025fast}.
Bridging this gap requires both algorithmic compression and hardware-aware system design.
Quantization, converting high-precision floating-point parameters and activations into compact integer formats,
has been shown to maintain model accuracy at 4--8-bit precision
while significantly reducing memory footprint, bandwidth requirements, 
and energy consumption per token~\cite{frantar2022gptq,xiao2023smoothquant,lin2024awq,xu2024empowering,hooper2024kvquant,dong2024qaq,zhang2024qhitter}.
On server platforms, quantization delivers substantial speedups
by leveraging native low-bit execution paths such as Tensor Cores~\cite{wang2022qgtc,yu2023boost,lin2024qserve,ladder,zhao2024atom,song2024powerinfer}.
In contrast, mobile GPUs often lack native support for low-bit arithmetic.
As shown in Table~\ref{tab:nn-performance-comparison}, INT8 inference on a Snapdragon~8 Gen~3 GPU is up to 2$\times$ \emph{slower} than FP16, while INT8 on the CPU (with native low-bit instructions) of the same SoC is 2$\times$ faster.
Without dedicated integer execution instructions,
every low-bit operation requires type promotion and conversion:
unpacking eight INT4 values from a 32-bit word costs approximately 16 shift and 16 mask operations,
exceeding the 8 multiply and 7 add operations of the dot-product itself (more details in Section~\ref{sec:empirical-study}).

Beyond these general DNN considerations, LLMs face a second bottleneck on mobile:
the dynamically growing KV cache.
Mobile GPUs organize memory as 2D textures with RGBA vector lanes~\cite{niu2024smartmem,zhang2021nn},
and the optimal data layout depends on the current decoding sequence length.
Existing frameworks~\cite{t-mac,t-man,bitnet} apply a fixed layout throughout decoding,
leading to misaligned memory accesses and periodic latency spikes as the KV cache grows.%
T-MAC~\cite{t-mac} and BitNet.cpp~\cite{bitnet} target CPU table-lookup computation,
which produces uncoalesced accesses on GPUs and cannot exploit the 2D texture memory hierarchy.
T-MAN~\cite{t-man} relies on NPU-specific table-lookup units and static data layouts
that cannot adapt to the dynamically growing KV cache during decoding.
Efficient quantized execution on mobile GPUs thus remains an open problem,
particularly when the optimal layout shifts at runtime with changing sequence length.

We  propose \projectname, a mobile GPU inference engine that realizes
this formulation through runtime layout adaptation, hardware-aligned packing, and
conversion-hiding integer execution.
We target 4-/8-bit group-wise quantization, 
which prior work~\cite{lin2024qserve, dong2024qaq} has shown to achieve best balance of efficiency and accuracy, 
rather than aggressive formats (e.g., 2-bit)~\cite{bitnet,chee2023quip,chen2024db} 
that often compromise accuracy.
Although layout selection~\cite{hu2026m2xfp}, %
low-bit packing~\cite{t-mac}, 
and kernel pipelining~\cite{cowan2020automatic} have each been studied,  
prior approaches generally optimize them independently or assume fixed execution configurations~\cite{t-man}. 
In contrast, we formulate mobile-GPU low-bit decoding as a dynamic,
quantization-aware layout-selection problem
and use a lightweight Layout Efficiency Proxy (LEP)
to guide runtime layout and packing choices
as the KV cache grows.
Our contributions are as follows:
\begin{itemize}[leftmargin=*,noitemsep,nolistsep]
    \item An empirical analysis showing that layout mismatch in the growing KV cache
        and data type conversion overhead are the key bottlenecks causing low-bit quantization
        to underperform on mobile GPUs.
    \item {A novel LEP-guided framework that combines quantization with a layout-aware runtime,
        enabling efficient low-bit LLM execution on mobile GPUs.}
    \item A set of novel mechanisms, including adaptive layouts for KV cache, hardware-aligned value packing,
        and an integer-native pipeline that defers type conversion outside the compute hot path.
    \item Extensive evaluation showing \projectname outperforms leading frameworks across multiple models and mobile GPUs. 
\end{itemize}

\projectname is evaluated on four LLMs with {\it grouped-query attention (GQA)}
and {\it traditional multi-head attention (MHA)},
namely Llama-3.2~\cite{dubey2024llama}, Qwen3~\cite{yang2025qwen3},
Gemma-2~\cite{team2024gemma}, and Phi-2~\cite{javaheripi2023phi},
under multiple precisions (FP16, W8A8, W4A8)%
\footnote{FP16: 16-bit floating-point for weight and KV cache. W8A8: 8-bit weights and 8-bit KV cache. W4A8: 4-bit weights and 8-bit KV cache.},
context lengths, and mobile GPU platforms.
\projectname achieves average speedups of 1.78$\times$ over LLAMA.CPP~\cite{llamacpp},
1.73$\times$ over MNN-LLM~\cite{wang2024mnn}, 1.58$\times$ over LiteRT-LM~\cite{litert-lm,TensorFlow-Lite}, 
and 1.47$\times$ over MLC-LLM~\cite{mlc-llm,chen2018tvm},
while eliminating the latency spikes observed in other frameworks during KV cache growth.
Average per-token energy consumption is reduced by 56--76\% relative to the state-of-the-art baselines.
\projectname enables stable execution of long-context models on resource-constrained devices where other frameworks exhibit severe degradation.

\section{LLM and Quantization Background}
\label{sec:quantization-background}

\subsection{Computation Flow in Transformer Decoder} 
\label{sec:background-computation-flow}

An autoregressive LLM generates tokens sequentially, 
with each step depending on all previously produced tokens. 
At step $N$, the model processes the current input embedding while appending new projections 
to a growing KV cache. 
This cache helps avoid  recomputation but its growing size  introduces challenges to 
relate to total memory and memory access costs/efficiency. 

\begin{figure}[!t]
	\centering
	\includegraphics[width=0.95\columnwidth]{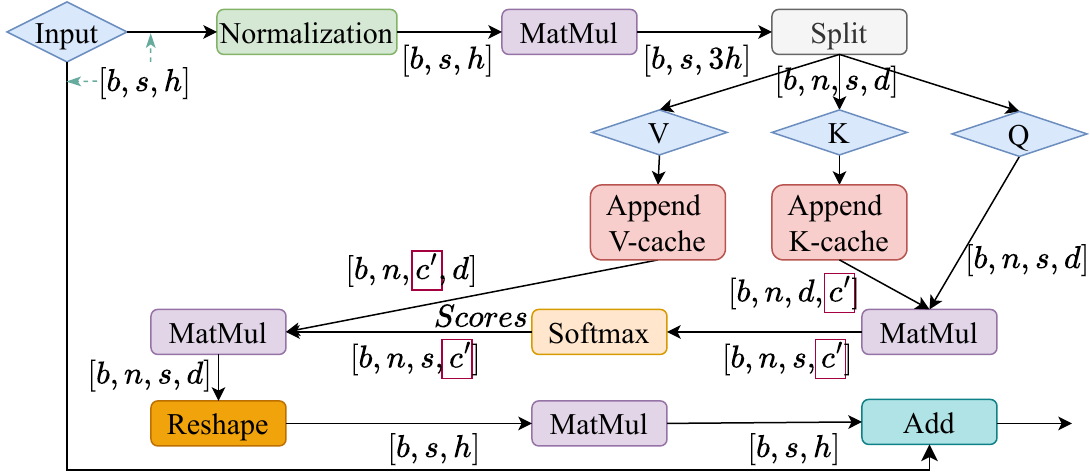}
	\caption{{\bf Computational graph of an autoregressive transformer decoder.}
	        Each new query ($Q$) attends to all cached keys ({\em K-cache}) and values ({\em V-cache}),
          causing the KV cache to grow and dynamically alter tensor shapes.}
	\label{fig:transformer-kvcache}
	\Description{Flow diagram of a decoder layer with Q, K, and V branches; the K-cache and V-cache append steps introduce a context-length dimension that grows with each decoded token.}
\end{figure}

Figure~\ref{fig:transformer-kvcache} illustrates the computation graph of 
a Transformer decoder layer during autoregressive decoding. 
Let $b$ denote the batch size and $s$ the number of tokens processed 
at the current step (typically $s{=}1$). 
Let $c^{\prime}$ be the updated {\em cache} length ($c^{\prime} = c_{\text{prev}} + s$). 
For the model, $n$ is the {\em number of attention heads}, 
$d$ is the {\em number of dimensions per head},
and $h = n \cdot d$ is the {\em hidden size}. 
The input embedding has shape $[b, s, h]$. 
After normalization and a linear projection, the representation becomes $[b, s, 3h]$, 
which is split into queries ($Q$), keys ($K$), and values ($V$), and reshaped into
\begin{equation}
Q, K, V \in \mathbb{R}^{[b, n, s, d]}.
\end{equation}
The new keys and values are appended to the caches,
\begin{equation}
K_{\text{cache}} \in \mathbb{R}^{[b, n, d, c^{\prime}]}, \qquad 
V_{\text{cache}} \in \mathbb{R}^{[b, n, c^{\prime}, d]},
\end{equation}

\begin{figure}[!t]
	\centering
	\includegraphics[width=0.92\columnwidth]{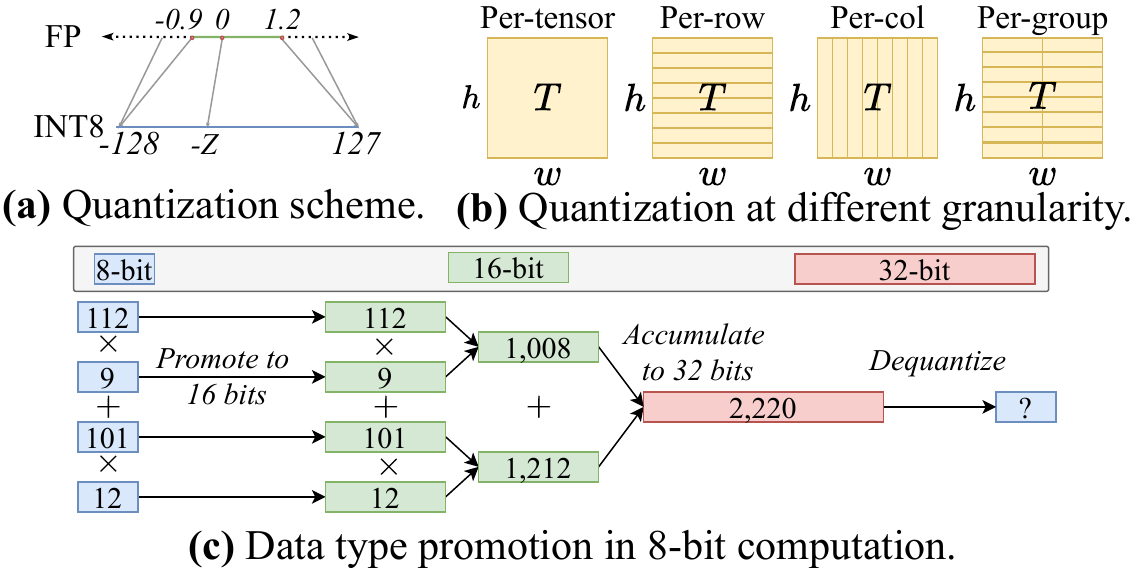}
	\caption{{\bf Overview of quantization techniques.}  
	      (a) Quantization scheme.
	      (b) Quantization at different granularities.  
	      (c) Type conversion pipeline for INT8 computation, involving promotion to INT16/INT32 
          and dequantization back.}
	\label{fig:quantization-overview}
	\Description{Three panels: mapping a floating-point range to INT8, four shading patterns for per-tensor, per-row, per-column, and per-group granularity, and an 8-bit multiply promoted to 16 bits, accumulated in 32 bits, and dequantized.}
\end{figure}

The memory layouts for these caches are discussed later; broadly, they are chosen to avoid transpose
operations in the subsequent attention computation.
The attention mechanism then computes similarity scores between the current queries and the cached keys:
\begin{equation}
{Scores} = QK^\top \in \mathbb{R}^{[b, n, s, c^{\prime}]}.
\end{equation}
After applying a softmax along the $c^{\prime}$ dimension, 
the result is multiplied by the value cache to obtain the context tensor
\begin{equation}
S = \text{Softmax}(Scores),
\qquad
SV \in \mathbb{R}^{[b, n, s, d]}.
\end{equation}
This result is reshaped back to $[b, s, h]$, passed through an output projection, and combined with the residual connection to form the block output $[b, s, h]$. In the final decoder layer, the output is projected into the vocabulary space to produce {\em logits} for next-token prediction.

While $s$ remains small and fixed in decoding, the cache length $c^{\prime} $ grows linearly with the number of generated tokens. 
This growth causes the attention computations to increase linearly and introduces different shapes of matrix multiplication. 
Thus, although the KV cache enables efficient reuse of past computations, its expansion also inflates memory consumption and complicates GPU execution optimization strategies such as static tiling and memory coalescing. 

\subsection{LLM Quantization}  
Quantization maps high-precision floating-point values into discrete and compact integer domains, 
reducing model size and lowering arithmetic cost.  
A real-valued tensor $x$ can be quantized using a uniform affine mapping 
with floating-point scale $s > 0$ and integer zero-point $z \in \mathbb{Z}$:
\begin{equation}
q = \operatorname{clip}\left(\mathrm{round}(x/s) + z,\; q_{\min}, q_{\max}\right),
\qquad \hat{x} \approx s \cdot (q - z),
\end{equation}
where $q$ is an integer within the target bit-width.  
As shown in Figure~\ref{fig:quantization-overview}(a), 
this affine mapping projects floating-point values into a bounded integer range (i.e., -128 to 127).

To balance accuracy and efficiency, 
various quantization schemes have been proposed. They also differ in the
granularity at which the scale and zero-point are chosen. 
Figure~\ref{fig:quantization-overview}(b) illustrates some of these different approaches.
\emph{Per-tensor quantization} applies a single scale across the entire tensor, 
offering low overhead but suffering from sensitivity to outliers~\cite{dong2024qaq}.
\emph{Per-row/column quantization} assigns independent scales per row or column, 
improving accuracy for tensors with heterogeneous distributions.  
\emph{Per-group quantization} generalizes this idea by partitioning tensors into groups of fixed size, 
balancing error reduction and implementation cost. 
Typical group sizes are 32 to 128~\cite{lin2024awq}.

\begin{figure}[!t]
	\centering
	\includegraphics[width=0.90\columnwidth]{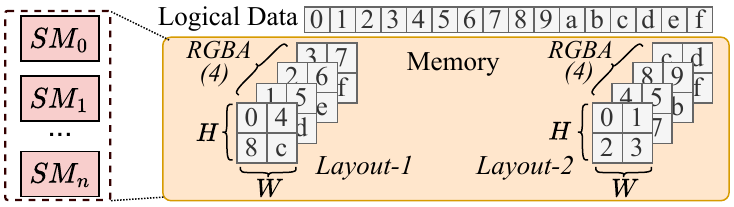}
	\caption{{\bf Illustration of 2D texture memory layout.}}
	\label{fig:gpu-texture-layout} 
	\Description{A 16-element logical array mapped onto GPU 2D texture memory in two different layouts, each drawn as four stacked RGBA channel planes.}
\end{figure}

\begin{figure}[!t]
	\centering
	\includegraphics[width=0.91\columnwidth]{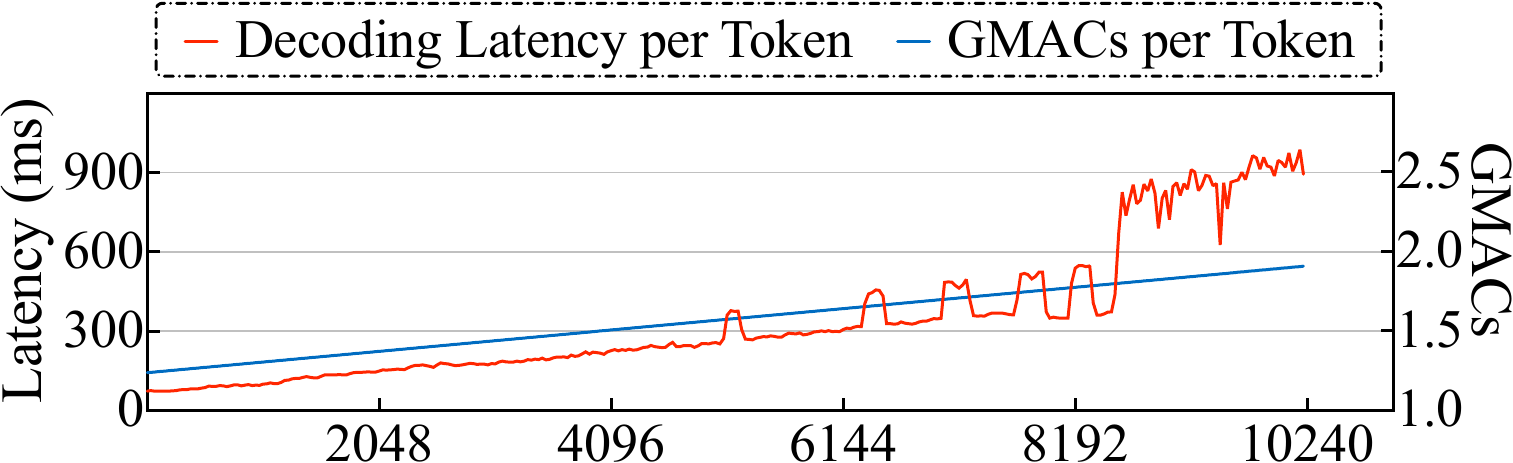}
	\caption{{\bf Decoding latency and computational workload per token as functions of decode sequence index.}}
	\label{fig:token-grows} 
	\Description{Line chart over 10240 decoded tokens: per-token latency rises in stair-steps from about 80 ms to over 900 ms, while GMACs per token increase smoothly and linearly.}
\end{figure}

\subsection{Constraints of Mobile GPUs}
As shown in Figure~\ref{fig:gpu-texture-layout}, the memory subsystem of mobile GPUs 
can be viewed as a 2-dimensional texture space~\cite{niu2024smartmem}:
a 2D grid with dimensions width $W$ and height $H$,
where each point stores a fixed-width vector -- typically four elements, like an RGBA pixel.
This structure, optimized for workloads with inherent 2D spatial locality,
allows the dedicated cache to efficiently fetch and cache data in 2D blocks
-- a fundamental divergence from the linear memory space in CPUs.
However, the ability to benefit from this feature depends on data layouts that align with the 2D memory space,
as misalignment leads to mandatory padding, uncoalesced accesses, and bandwidth under-utilization. The optimal layout is generally dictated by the specific computation being performed. 
As an illustration of possibilities, 
Figure~\ref{fig:gpu-texture-layout} also illustrates two distinct data layouts for a tensor of length 16,
where Layout-1 uses row-major ordering, and Layout-2 leverages column-major ordering.

On server-class platforms, quantization is effectively supported through optimized
low-bit instructions  such as those seen in AVX-512, VNNI, and CUDA-INT4/INT8 extensions.
However, mobile GPUs are primarily designed for FP16/FP32 pipelines. 
Figure~\ref{fig:quantization-overview}(c) shows the sequence of data type conversion
required for INT8 operands in the absence of specialized instructions: 
(i) unpacking,
(ii) promotion to INT16 for multiplication,
(iii) accumulation into INT32,
and (iv) dequantization back to FP32 for subsequent operations.
These steps add significant computational cost and register pressure.

\begin{table}[!t]
	\small
	\setlength{\tabcolsep}{6pt}
	\caption{{\bf Comparison of LLM decoding latency on mobile platforms with FP16 versus low-bit representation.}} 
	\label{tab:nn-performance-comparison}
	\begin{tabular}{l|r|R{1.0cm}|R{1.0cm}|r}
		\toprule
		\multirow{2}{*}{Model}        & Data     & \multicolumn{2}{c|}{Latency per Token} & GPU \\ \cline{3-4}
		~                                                       & Type     & CPU     & GPU     & Speedup \\   \midrule
		\multirow{2}{*}{TinyLLama-1B~\cite{zhang2024tinyllama}} & FP16     & 61.9ms  & 56ms    & 1.11$\times$ \\
		~                                                       & INT8     & 31.2ms  & 67.8ms  & {\bf 0.46$\times$} \\ \midrule
		\multirow{2}{*}{Qwen3-1.7B~\cite{yang2025qwen3}}        & FP16     & 78.8ms  & 73.6ms  & 1.07$\times$ \\
		~                                                       & INT8     & 45.4ms  & 130.2ms & {\bf 0.35$\times$} \\ \midrule
		\multirow{2}{*}{Llama-3.2-3B~\cite{dubey2024llama}}     & FP16     & 140.5ms & 137.4ms & 1.02$\times$ \\
		~                                                       & INT8     & 80.1ms  & 159.2ms & {\bf 0.50$\times$} \\
		\bottomrule
	\end{tabular}
\end{table}

\section{Empirical Study on LLM Inference}
\label{sec:empirical-study}

To understand the performance impact of the steps outlined above, 
we profile the decoding latency of various LLMs on a commercial mobile device (OnePlus 12) with MNN-LLM~\cite{wang2024mnn},
comparing FP16 and INT8 on both CPU and GPU. 
The results, summarized in Table~\ref{tab:nn-performance-comparison}, 
first confirm the expected pattern for FP16 execution: 
GPUs consistently outperform CPUs, 
leveraging their superior floating-point throughput. 
However, INT8 quantization reverses this trend. 
INT8 reliably accelerates inference on the CPU (i.e., as compared to FP16), 
by leveraging reduced memory bandwidth and efficient integer arithmetic. 
However, on the mobile GPU, INT8 consistently results in degraded performance,
with execution that is even slower than the CPU's INT8 path.
This inversion stems from the overhead of type promotion and bit-level manipulation
that low-bit computation requires on mobile GPUs --
operations  that existing frameworks do not handle efficiently.

The dynamically growing KV cache introduces a second bottleneck in the memory subsystem.
Figure~\ref{fig:token-grows} highlights the impact of KV cache size on latency. 
The blue curve shows that the arithmetic workload (measured in GMACs per token) 
increases in a consistent linear fashion with sequence length. 
In contrast, the red curve reveals that decoding latency per token grows much faster and exhibits frequent stair-step spikes.
These spikes occur when changes in KV tensor size and shape cause the existing layout to become
suboptimal.
\par

\section{System Design}\label{sec:design}

{
The key idea of \projectname is to make low-bit LLM execution layout-adaptive on mobile GPUs.
The motivation for this approach comes from the fact 
that layout and unpacking costs interact on the critical path, 
optimizing one in isolation yields limited gains.
}
More specifically, we first formalize the notion of the Layout Efficiency Proxy (LEP) (Section~\ref{sec:design:formulation}), 
which models the memory and data unpacking costs that dominate the runtime. 
Grounded in this formalization, we introduce three complementary mechanisms: 
i) adaptive layouts that evolve dynamically with sequence length (Section~\ref{sec:design:layout}), 
ii) quantization-aware packing designed to minimize unpacking overhead (Section~\ref{sec:design:packing}), 
and iii) a type {conversion-hiding} computation pipeline that preserves these gains through low-bit integer execution (Section~\ref{sec:design:pipeline}).
We use $H$ (attention heads), $D$ (hidden dimension per head), and $T$ (current sequence length) for the transformer dimensions from Section~\ref{sec:background-computation-flow}.

\subsection{Optimization Overview}\label{sec:design:formulation}

\begin{figure}[!t]
	\centering
	\includegraphics[width=.9\columnwidth]{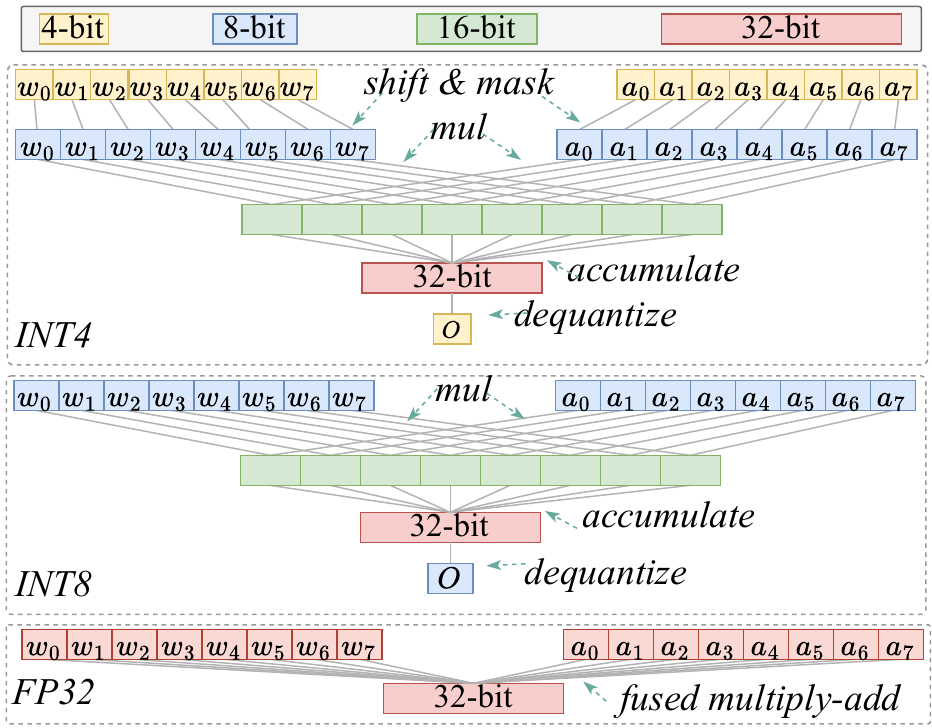}
	\caption{
    \textbf{Execution pipelines for a length-8 dot product with INT4, INT8, and FP32 formats.} 
		FP32 uses direct fused multiply-add ({\tt fma}); INT8 performs integer {\tt mul} with 32-bit {\tt accumulation}; INT4 additionally unpacks sub-byte weights with {\tt shift} and {\tt mask} instructions.
    }

	\label{fig:quant_computation_pipeline} 
	\Description{Three panels for INT4, INT8, and FP32 dot products; INT4 adds a shift-and-mask unpack step, INT8 multiplies and accumulates in 32 bits, and FP32 uses a single fused multiply-add.}
\end{figure}

Two cost factors jointly determine runtime performance and guide our optimizations.
The first is \emph{data access cost}.
Different attention operators favor different stride-1 axes:
$QK^{T}$ benefits from $D$-contiguous layouts for stride-1 query access,
while $SV$ benefits from $T$-contiguous layouts for efficient token traversal.
A misaligned layout forces strided fetches and under-utilizes the GPU's RGBA vector lanes.
The data access cost, in summary,  depends on the layout configuration and further changes with sequence length.
The second factor is \emph{unpacking overhead}.
As shown in Figure~\ref{fig:quant_computation_pipeline},
an FP32 8-element dot product uses 8 {\tt FMA} instructions, 
but low-bit values must be unpacked into byte- or word-aligned operands
before the mobile GPU can perform arithmetic on them.
INT8 values are already byte-aligned, requiring only 8 {\tt mul} and 7 {\tt add} for accumulation.
INT4 values, however, are sub-byte: eight 4-bit values share a single 32-bit word, and extracting them
requires roughly 16 {\tt shift} and 16 {\tt mask} operations
on top of the 8 {\tt mul} and 7 {\tt add},
which can rival or exceed the arithmetic cost itself.
Because this overhead depends on the packing scheme,
it must be modeled explicitly rather than folded into a generic compute cost.

To guide our optimization decisions, 
we define the Layout Efficiency Proxy (LEP) for a layout $\mathcal{L}$ of tensor $t$. 
Here, a layout configuration $\mathcal{L} = (\pi, \rho, \phi)$ specifies three choices:
\begin{itemize}[leftmargin=*,noitemsep,nolistsep]
    \item $\pi$ (contiguous-axis order) determines which logical tensor axis is stride-1 in memory, such as head $H$, hidden dimension $D$, or token $T$.
    \item $\rho$ (2D texture placement) determines how the ordered tensor axes are mapped to texture rows and columns on the mobile GPU.
    \item $\phi$ (low-bit value packing) determines how 4-/8-bit values are arranged within a machine word or RGBA lanes.
\end{itemize}

The LEP is formulated as a weighted sum of two dominant cost factors:
\begin{equation}
\mathrm{LEP}(\mathcal{L}, t) =
\alpha_1 \cdot \frac{\mathrm{Bytes}(\pi, \rho, \phi, t)}{\mathrm{BW}_{\text{tex}}}
+ \alpha_2 \cdot \mathrm{UnpackOps}(\phi)
\label{eq:lep}
\end{equation}

\noindent The key is that both LEP components can be efficiently computed by combining analytical modeling with static analysis.
\begin{itemize}[leftmargin=*,noitemsep,nolistsep]
    \item {Data access cost ($\mathrm{Bytes}(\cdot)/\mathrm{BW}_{\text{tex}}$)}: 
    this term is derived \textit{analytically} by modeling the tensor ($t$) access patterns
    against the memory mapping defined by $\pi$ and $\rho$. 
    For an operator and  $t$, we simulate the memory traversal pattern 
    to count the bytes moved between DRAM and on-chip cache, 
    considering cache line sizes and bank conflicts. 
    
    \item {Unpacking cost ($\mathrm{UnpackOps}(\phi)$)}: 
    this term is quantified through \textit{static code analysis}.
    For each quantization packing scheme $\phi$,
    we generate the corresponding kernel code and count the number of integer arithmetic 
    and bitwise operations (shifts, masks, AND/OR) required to unpack a single packed word.
    This operation count is normalized to represent the overhead per arithmetic operation.
    
    \item {Weight calibration ($\alpha_1$, $\alpha_2$)}:
    $\alpha_1$ and $\alpha_2$ are device-specific normalization weights 
    calibrated once per GPU
    through offline microbenchmarks ($<$10 minutes) that separately stress
    the memory subsystem and integer ALU.
    The weights are model-independent and remain fixed 
    across all models and sequence lengths.
    On our evaluated GPUs, $\alpha_1 = 0.5$ and $\alpha_2 = 0.7$ yield the best layout selection accuracy.
\end{itemize}

To make LEP evaluation efficient at runtime,
we first pre-compute $\mathrm{Bytes}(\pi, \rho, \phi, t)$ and $\mathrm{UnpackOps}(\phi)$ 
for all supported layout configurations during an offline profiling phase.
During runtime selection, \projectname queries these existing values 
and performs the weighted sum in Equation~\ref{eq:lep}.
This hybrid analytical-static approach allows \projectname to efficiently 
evaluate thousands of $(\mathcal{L}, t)$ combinations with minimal overhead. 
The LEP serves as a comparative metric to identify the most efficient layout configuration for the dominant operators in transformers (i.e., {\tt MatMul}).
\begin{figure}[t!]
  \centering
  \includegraphics[width=0.85\columnwidth]{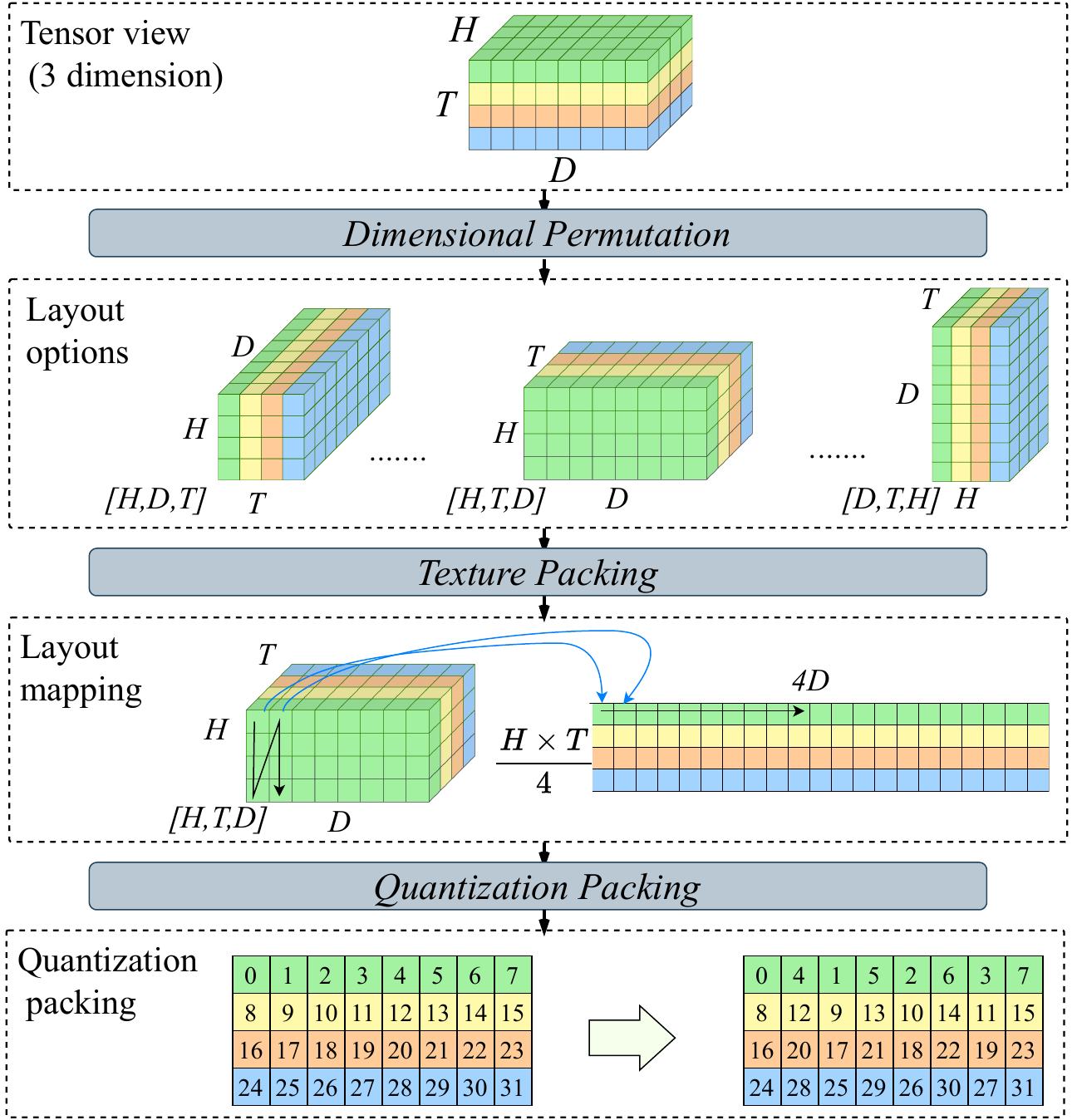}
  \vspace{-0.5em}
  \Description{Four-stage flowchart: a 3D tensor is permuted into candidate layouts, packed into a 2D texture, and its elements reordered into an interleaved quantization packing.}
  \caption{\textbf{Layout optimization pipeline.}}
  \vspace{-1.0em}
  \label{fig:layout-overview}
\end{figure}

Figure~\ref{fig:layout-overview} shows an example of the layout optimization pipeline for tensor $K$ with shape $[H,D,T]$ under configuration $\mathcal{L}$. 
First, the dimension $D$ is chosen as the stride-1 axis since computation is performed along $D$, ensuring consecutive cache access. 
The tensor is then mapped into 2D texture space along the $H$ dimension, 
and finally, the low-bit quantized values are rearranged and packed according to $\phi$ for efficient dequantization.
The subsequent sections describe our detailed mechanisms for minimizing the LEP. 

\subsection{Adaptive Layouts for Decoding}\label{sec:design:layout}
Having established the LEP as our efficiency metric,
we now present the adaptive layout mechanism that minimizes this proxy.
The challenge translates to dynamically selecting, 
for each KV cache operation,
the layout configuration $\mathcal{L} = (\pi, \rho, \phi)$ that
yields the lowest LEP given the current tensor $t$ and KV cache.

\compactparagraph{Layout adaptation with sequence length.}
This section addresses the $\mathrm{Bytes}/\mathrm{BW}_{\text{tex}}$ term through runtime adaptation.
Specifically, in layout adaptation with sequence length, 
the optimal stride-1 axis for the KV cache is a function of sequence length. 
{\em Short contexts} (e.g., $< 256$) are dominated by the $QK^\top$ operation, favoring a $D$-contiguous layout. 
{\em Long contexts} (e.g., $> 512$) are dominated by $SV$, 
favoring a $T$-contiguous layout. 
Choosing a static layout leads to inefficient memory access and severe latency spikes (Figure~\ref{fig:token-grows}).

We adopt a greedy layout evolution strategy.
Algorithm~\ref{alg:layout-evolution} formalizes this procedure,
which periodically evaluates candidate layouts and selects the one
minimizing LEP for the current and near-future sequence lengths.
The algorithm generates a set of candidate layouts $\mathcal{C} = \{(\pi, \rho, \phi)\}$
by enumerating permissible stride-1 axis permutations (e.g., $D$-major vs.\ $T$-major)
along with corresponding texture packing schemes $\rho$ that comply with the mobile GPU's memory constraints.
It evaluates each candidate by computing its LEP score at the current sequence length $t$,
estimating the efficiency of the predominant operator (e.g., $QK^\top$ or $SV$) under that layout.
Because LEP uses precomputed values and one weighted sum per candidate,
selection completes in microseconds with negligible decoding overhead.
The lowest-LEP candidate is adopted only if it measurably improves
upon the current configuration; otherwise, the current layout is retained.

\begin{algorithm}[t]
\DontPrintSemicolon
\small
\SetNoFillComment
\caption{Greedy Layout Evolution}
\label{alg:layout-evolution}
\SetKwInput{INPUT}{Input: }
\SetKwInput{OUTPUT}{Output: }
\SetKw{Return}{return}
\SetKwRepeat{Do}{do}{while}%

\INPUT{Current layout $\mathcal{L}_{current}$, sequence length $t$, cache dimensions $(H, D, T)$}
\OUTPUT{New layout $\mathcal{L}_{new}$}

$\mathcal{C} \gets \text{GenerateCandidates}(H, D, T)$

\ForEach{candidate $\mathcal{L}_c \in \mathcal{C}$ }{
	$ \text{LEP}_c \gets \alpha_1 \cdot \frac{\text{Bytes}(\mathcal{L}_c, t)}{\text{BW}_{\text{tex}}} + \alpha_2 \cdot \text{UnpackOps}(\mathcal{L}_c) $
}

$\mathcal{L}_{best} \gets \arg \min_{\mathcal{L}_c \in \mathcal{C}} \text{LEP}_c$

\If{$\text{LEP}_{best} < \text{LEP}_{current}$}{
	$\quad \mathcal{L}_{new} \gets \mathcal{L}_{best}$
}\Else {
	$\quad \mathcal{L}_{new} \gets \mathcal{L}_{current}$
}
\end{algorithm} 

\compactparagraph{Folding along non-reduction axes.}
Mobile GPUs impose strict limits on the width and height of texture memory 
(e.g., $\leq$ 16K pixels on each dimension).
As the KV cache grows in long context, tensor dimensions often exceed these limits, 
making direct mapping into 2D textures infeasible. 
A naive folding strategy partitions oversized dimensions arbitrarily. 
However, if folding occurs along a {\em reduction axis} -- 
e.g., folding the hidden dimension $D$ in $QK^\top$ 
or folding the token dimension $T$ in $SV$ -- the reduction spans across multiple tiles. 
This introduces strided, cross-cache-tile fetches, 
increasing memory traffic and degrading cache locality.

\begin{figure}[t!]
  \centering
  \includegraphics[width=0.82\columnwidth]{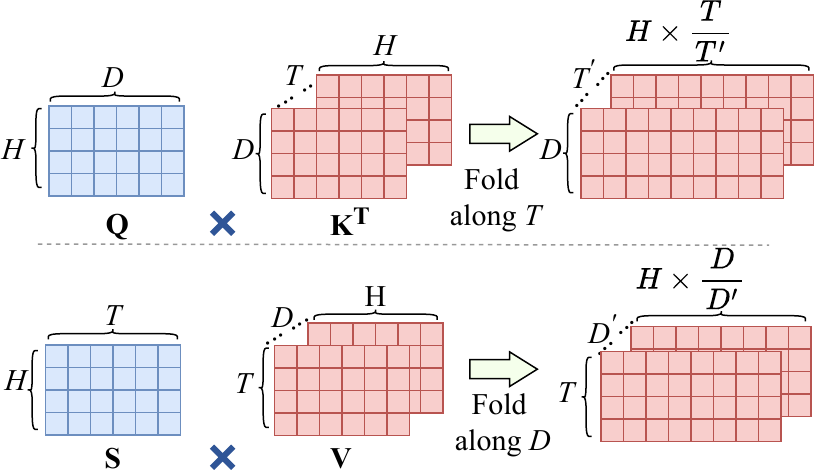}
  \vspace{-0.5em}
  \caption{\textbf{Two folding examples.}}
  \vspace{-0.5em}
  \label{fig:folding-example}
  \Description{Two matrix multiplication diagrams showing the K transpose operand folded along T and the V operand folded along D into narrower stacked tiles.}
\end{figure}

We adopt an operator-aware folding policy.
The key insight is that {\it folding should preserve contiguous access along the dimension
that is reduced during a computation-intensive operation (e.g., {\tt MatMul})},
as it is critical for achieving a stride-1 memory access pattern
and high memory bandwidth utilization.
Consequently, whenever folding is necessary, it is applied exclusively along non-reduction axes.
This ensures the reduction dimension remains contiguous, preserving stride-1 access.
As shown in Figure~\ref{fig:folding-example}, for $QK^\top$ (reduction over $D$), we fold along $T$;
and for $SV$ (reduction over $T$), we fold along $D$.
This principle extends to other operators like {\tt Softmax},
where reduction occurs along the last dimension, so we fold along preceding dimensions.
This policy leverages operator semantics to make a locally optimal choice with zero search overhead.

\compactparagraph{Append-only segments and phase-aware transitions.}
As the sequence length increases, the initially optimal KV cache layout can become suboptimal. 
Conventional systems typically resort to one of two costly approaches: 
1) performing a re-layout of the entire cache -- a process 
that copies hundreds of megabytes of data and imposes significant overhead, 
or 2) leaving the layout unchanged and accepting suboptimal performance.
Both result in sharp latency spikes.
A naive alternative is to switch layouts greedily at every small improvement 
in the Layout Efficiency Proxy (LEP). 
While this avoids re-layout, it produces many small fragments of the cache, 
significantly increasing iterator overhead and complicating memory management.

We propose an append-only segmentation strategy coupled with phase-aware transitions.  
The key idea is {\em Append-only Segments}, where  KV cache is divided into immutable segments. 
Each segment records its layout $(\pi,\rho,\phi)$ and kernel schedule parameters (e.g., unroll factor, block size). 
When LEP indicates that a more efficient configuration is available, 
new tokens are appended into a new segment configured with that new layout. 
Previous segments remain intact, eliminating the need for copying or re-layout. 
To prevent excessive fragmentation, we design {\em phase-aware transitions}.
Specifically, we apply a heuristic to LEP evaluation and trigger layout switches
only when improvements persist over a sufficient sequence length. 
This compresses short-term fluctuations into a few stable computational phases, often aligned with operator-level shifts, such as transitions from \(QK^\top\)-heavy to \(SV\)-heavy workloads.
{
This design avoids copying or relayout during transitions;
switching only initializes metadata for a new segment.
Across our 10K-token runs, it creates 2.25 segments on average
and at most five.
Because each segment is contiguous,
the runtime traverses only a small number of segments, 
and total segmentation overhead remains below 1\% of end-to-end decoding latency.
}

\subsection{Quantization-Aware Packing}\label{sec:design:packing}
Low-bit quantization reduces memory footprint but introduces significant unpacking overhead
when data formats are misaligned with hardware execution units.
Modern mobile GPUs expose native integer dot-product instructions (e.g., {\tt dp4a} and {\tt udot8}),
which multiply multiple INT8 pairs and accumulate the result into an INT32 register in a single cycle.
To exploit these primitives effectively, operands must be packed into specific 32-bit vector layouts.

Naively packing low-bit values (e.g., INT4) contiguously into a 32-bit word misaligns them with byte-addressable hardware. Eight linearly packed INT4 values cross byte boundaries, forcing the GPU to extract each value and expand it to INT8 before integer arithmetic.
As quantified in Section~\ref{sec:design:formulation}, 
this expansion can cost more than the computation itself.

\begin{figure}[!t]
	\centering
	\includegraphics[width=0.91\columnwidth]{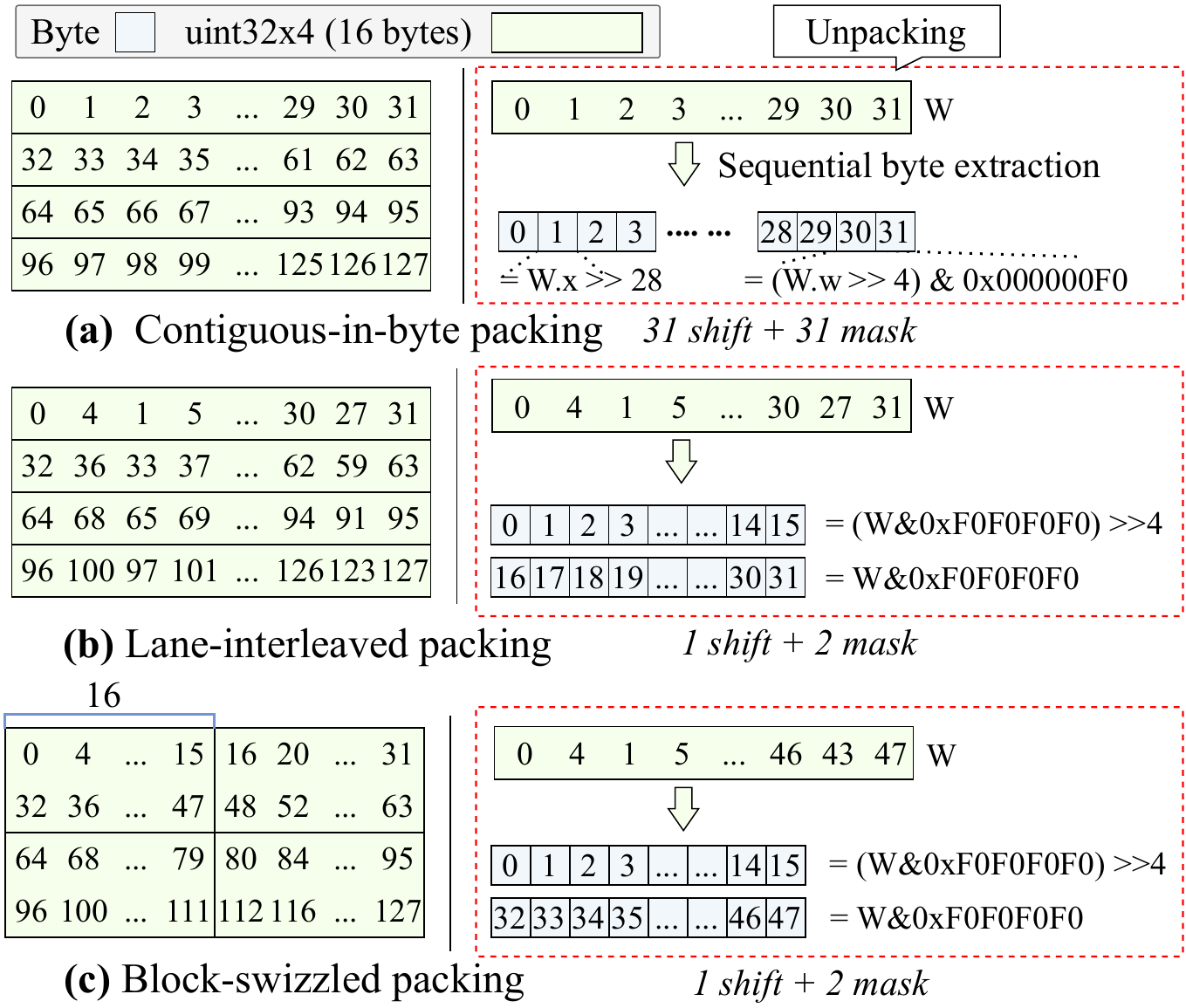}
	\caption{{\bf Different quantization packing schemes.}}
	\label{fig:packing-scheme} 
	\Description{Three byte-index grids with their unpacking formulas: contiguous packing needs 31 shifts and 31 masks, while lane-interleaved and block-swizzled packing need 1 shift and 2 masks.}
\end{figure}

To mitigate this inefficiency, we treat quantization packing 
as a schedulable dimension and redesign it to minimize the {\tt Unpacking} term in LEP. 
Specifically, we introduce a set of structured packing families.
Figure~\ref{fig:packing-scheme} illustrates the three schemes.
The {\em contiguous-in-byte scheme} maximizes storage density
but requires about 31 shift and 31 mask operations to extract values from a four-word vector, the highest unpack cost.
{\em Lane-interleaved packing} rearranges nibbles so that each byte holds two aligned INT4 values,
matching the four RGBA lanes of texture memory.
This reduces unpacking to a fixed sequence of one shift and two masks per word, independent of tensor shape.
{\em Block-swizzled packing} further organizes values into $2 \times 2$ texel tiles
aligned with texture cache blocks,
combining the constant-time unpacking of lane interleaving with improved memory coalescing.

Formally, if packed operands are expressed as

\vspace{-1.2em}
$$
A = \sum_{i=0}^{3} a_i \cdot 2^{8i}, \qquad 
B = \sum_{i=0}^{3} b_i \cdot 2^{8i}, \quad a_i,b_i \in \mathbb{Z}_8,
$$
the {\tt dp4a} instruction computes
\vspace{-0.6em}
$$
\mathrm{dp4a}(A,B,r_{\text{acc}}) = r_{\text{acc}} + \sum_{i=0}^{3}(a_i \times b_i), \quad r_{\text{acc}} \in \mathbb{Z}_{32}.
$$
Our packing strategies guarantee that unpacked INT4 values expand directly into the aligned INT8 operands required by such primitives, with only constant-time bit operations.

Compared to naive packing, where each word requires up to 20 unpacking operations, 
our reordered packing reduces the requirement to approximately three instructions. 
Microbenchmarks show that this reduces unpacking overhead by about 80\%
and increases integer pipeline utilization accordingly.
Without this, the instruction cost of unpacking can negate the memory savings from quantization.

\begin{table}[t!]
  \caption{8-bit dot-product primitives on different platforms.}
  \small
  \setlength{\tabcolsep}{5pt}
  \centering
  \begin{tabular}{ll}
    \toprule
    \textbf{Platform} & \textbf{Instruction} \\
    \midrule
    Adreno/MALI GPU & \texttt{udot8\_acc} / \texttt{dot8\_acc} \\
    AMD GPU (GCN/RDNA) & \texttt{v\_dot4\_u32\_u8} / \texttt{v\_dot4\_i32\_i8} \\
    Intel GPU (Xe) & \texttt{dpas.s32.s8.s8} \\
    NVIDIA GPU & \texttt{dp4a.s32.s8.s8} \\
    \bottomrule
  \end{tabular}
  \label{tab:dot8-instruction}
\end{table}

\compactparagraph{Discussion: applicability to other GPUs.}
Our implementation targets Adreno and Mali GPUs (Section~\ref{sec:portability}),
but the packing design requires only 8-bit integer dot-product accumulate instructions.
Table~\ref{tab:dot8-instruction} shows that similar primitives exist on AMD, Intel, and NVIDIA GPUs,
making the approach portable beyond mobile.
Extension to CPUs and NPUs is less straightforward:
low-bit systems on CPU/NPU~\cite{t-mac,bitnet,t-man} use table-lookup 
over SIMD lanes co-designed with linear memory hierarchies,
whereas our packing and layout co-design targets warp-parallel execution.
Comparison across these platforms is a direction for future work.

\subsection{Type Conversion-Hiding Computation}\label{sec:design:pipeline}
Conventional low-bit computation on mobile GPUs incurs repeated data type conversions.
INT8 multiplication requires upcasting to INT16 to avoid overflow,
and many frameworks further convert quantized tensors to FP16 for {\tt MatMul}.
These conversions account for over 30\% of arithmetic instructions in INT4 execution
and leave the integer dot-product units underutilized,
because conversion overhead sits directly on the critical path.

We introduce an integer-native, tile-based pipeline that eliminates conversions from the core computation path.
We reorganize the computation around tiles,
which are small and contiguous blocks of the weight or activation matrix
that serve as the fundamental unit of data transfer and parallel computation.
Our design is built on three key principles: 1) maintaining all hot-path computations in the integer domain,
2) deferring dequantization and floating-point conversion to an epilogue stage,
and 3) overlapping this epilogue asynchronously with the computation of subsequent tiles.
This represents a paradigm shift from {\em converting-then-computing} to {\em computing-then-converting}.

\begin{figure}[!t]
	\centering
	\includegraphics[width=0.90\columnwidth]{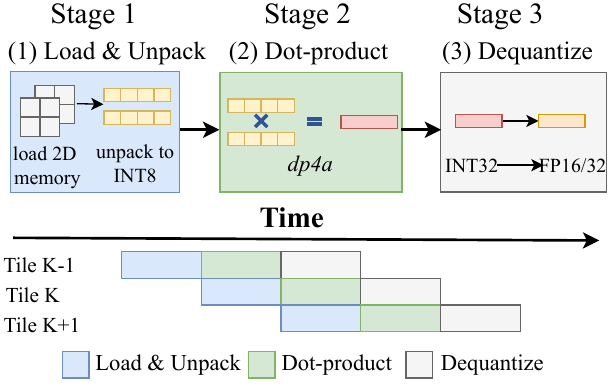}

	\caption{{\bf Tile-based pipelining for type {conversion-hiding} computation.}
          The pipeline consists of three stages: 
          (1) {\em Load \& Unpack}, 
          (2) {\em Integer Compute}, and 
          (3) {\em Deferred Conversion}. 
          Operations overlap across tiles, ensuring the hot path remains fully in the integer domain.}
	\label{fig:conversion-pipeline} 
  \vspace{-1em}
	\Description{Three pipeline stages, load and unpack, dp4a dot-product, and dequantize, with a timeline showing the stages of consecutive tiles overlapping in time.}
\end{figure}

As shown in Figure~\ref{fig:conversion-pipeline}, 
the pipeline is organized at the tile granularity and comprises three tightly overlapped stages.
First ({\em Load \& Unpack}), tiles are loaded in their compactly-packed form from texture memory. 
Leveraging the reordered packing scheme from the previous section, %
we unpack INT4 data into byte-aligned INT8 values using only $\approx$3 bitwise operations (1 {\tt shift} and 2 {\tt mask}) per 32-bit word -- 
a 5$\times$ improvement over the 16 operations needed for naive contiguous packing. 
This unpacking occurs uniformly in registers, 
eliminating thread divergence and minimizing $\mathrm{UnpackOps}$.

In the second stage ({\em Dot-product}), 
the unpacked, byte-aligned tiles are immediately fed into integer dot-product primitives (e.g., \texttt{dp4a}, \texttt{udot8}), 
which accumulate results directly into INT32 registers. 
This stage is entirely free of type conversions, 
allowing the integer units to operate at peak utilization. 
Tile sizes are co-designed with the adaptive layouts (Section~\ref{sec:design:layout}) 
to balance register pressure and memory coalescing, keeping traffic near the LEP lower bound.

In the third stage ({\em Dequantization}),
the integer accumulations are converted to FP16/FP32 and fused with dequantization.
This epilogue runs only at tile boundaries, asynchronously via warp specialization.
As shown in Figure~\ref{fig:conversion-pipeline},
warp group A computes on tile $k$ while group B loads tile $k{+}1$ and group C converts the results of tile $k{-}1$.
Synchronization is needed only at tile boundaries,
which effectively removes conversion latency from the critical path.
{
Except when parallelism or compute slack is limited, 
as in very short contexts (sequence lengths below 12 in our kernels) or small matrices, 
the pipeline effectively overlaps conversion with computation. 
Such cases are uncommon in our evaluated workloads. 
Even under these conditions, conversion remains outside the inner dot-product loop, 
and only a portion of the epilogue latency is exposed on the critical path.}

The conventional pipeline requires two conversions per tile
(unpacking before computation and dequantization after),
which creates register traffic and leaves the integer ALUs idle.
Our pipeline eliminates conversions from the hot path;
the single deferred dequantization per tile is hidden behind the next tile's computation.
In practice, this reduces conversion instructions by over an order of magnitude
and unpack overhead by $\approx$80\% (Section~\ref{sec:evaluation-breakdown}).

\section{Evaluation}
\label{sec:evaluation}

This section evaluates \projectname's effectiveness in terms of latency, energy consumption, memory/cache efficiency, optimization contributions, and portability across mobile GPUs.

\subsection{Experiment Setup}  

\begin{table}
	\small
	\setlength{\tabcolsep}{7pt}
	\caption{\textbf{Model perplexity on the WikiText2~\cite{merity2016pointer} dataset (lower is better) for different precisions.}}%
  \label{tab:eval-model-accuracy}
	\begin{tabular}{l|r|r|r|r}
		\toprule
		\multirow{2}{*}{Model}  & \multirow{2}{*}{\#Params}  & \multicolumn{3}{c}{Perplexity $\downarrow$} \\ \cline{3-5}
		~                       & ~      & FP16   & W8A8   & W4A8 \\   \midrule
		Llama-3.2-1B~\cite{dubey2024llama}  & 1.24B & 13.75 & 14.31 & 15.34 \\
		Qwen3-1.7B~\cite{yang2025qwen3}     & 1.72B & 9.84 & 9.97 & 10.28 \\
		Gemma-2-2B~\cite{team2024gemma}     & 2.61B & 18.09 & 20.33 & 21.40 \\
		Phi-2-2B~\cite{javaheripi2023phi}   & 2.78B & 17.41 & 18.15 & 19.94 \\
    \bottomrule
	\end{tabular}
\end{table}

\compactparagraph{Workloads.}
We select four representative LLMs that are suitable for mobile GPU deployment: 
{\em Llama-3.2-1B}~\cite{dubey2024llama}, {\em Qwen3-1.7B}~\cite{yang2025qwen3}, 
{\em Gemma-2-2B}~\cite{team2024gemma}, and {\em Phi-2-2B}~\cite{javaheripi2023phi}.
{
They cover three attention mechanisms: 
grouped-query attention (GQA) as in  Llama-3.2 and Qwen3, 
multi-head attention (MHA) as in Phi-2, 
and alternating local-global attention in Gemma-2.}
These models are widely used in on-device studies and fit within the 8--16GB 
unified memory (shared with the OS and other apps) of current mobile GPUs; 
larger models (e.g., 7--8B) remain impractical at full precision.
For each model, we evaluate two low-bit precision settings:
\emph{W8A8}, with 8-bit weights, activations, and KV cache under group-wise scaling;
and \emph{W4A8}, with 4-bit weights and 8-bit activations and KV cache ---
a balance between compression and accuracy validated by prior work~\cite{xiao2023smoothquant,lin2024qserve}.
The group size is 32, following~\cite{frantar2022gptq,lin2024awq}.
Although our main focus is low-bit quantization, 
we include {\em FP16} to show  our adaptive layouts benefit higher-precision computation as well.

Before focusing on latency, energy, and memory behavior in the rest of this section, 
we demonstrate that the quantization levels chosen here result in high accuracy. 
Table~\ref{tab:eval-model-accuracy} reports model-level accuracy on standard benchmarks. 
Across all four models, W8A8 incurs only a modest perplexity increase over FP16 (0.1--2.2 points), 
while W4A8 remains within 0.4--3.3 points of FP16, 
confirming that group-wise quantization at these precisions preserves model quality.

\projectname is independent of the quantization algorithm and focuses on execution efficiency.
Therefore, for a given quantization scheme, \projectname provides the same model accuracy as other frameworks.
The remainder of our evaluation therefore focuses on system-level performance.

\compactparagraph{Baselines.}
We compare \projectname with 4 widely used LLM inference engines on mobile GPUs:
LLAMA.CPP~\cite{llamacpp} (31914624), MNN-LLM~\cite{wang2024mnn} (7f1197e0), 
LiteRT-LM~\cite{litert-lm} (26f2178, powered by TensorFlow-Lite~\cite{TensorFlow-Lite}),
and MLC-LLM~\cite{mlc-llm} (8f49ea6b, powered by TVM~\cite{chen2018tvm}).
Several related frameworks, such as bitnet.cpp~\cite{bitnet} and
other works including 
Transformer-Lite~\cite{transformerlite2024},
Ladder~\cite{ladder}, 
PowerInfer-2~\cite{xue2024powerinfer},
and ExecuTorch~\cite{executorch}, 
do not support the tested models on mobile GPUs. 
This is because they lack operator implementations or otherwise have hardware dependencies. 
In the reported results, we enable the auto-tuning capabilities available
in all baselines to achieve the best possible performance. 
The batch size is set to 1 in mobile environments.
Energy consumption is collected through \textit{Android system interface}.
We run each experiment 50 times and report the average,
as the variance is negligible. 

\begin{figure}[t!]
  \centering
  \includegraphics[width=0.91\columnwidth]{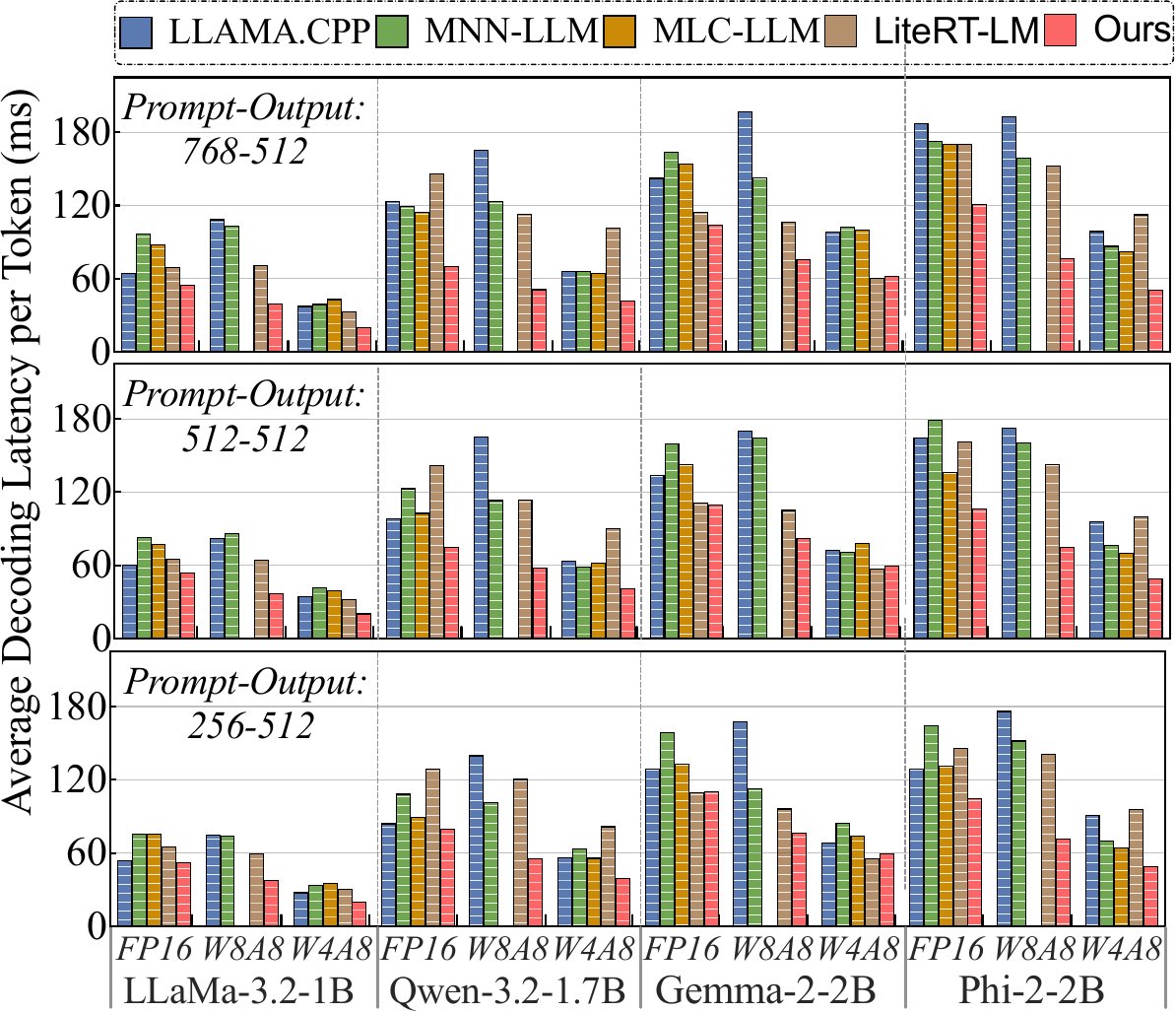}
  \caption{\textbf{Average latency per token comparison across models, precisions, and Prompt–Output ($P$–$O$) window.}}
  \label{fig:eval-e2e-seq-combined}
  \Description{Grouped bar charts of per-token latency for four models at FP16, W8A8, and W4A8 under three prompt-output settings; the Ours bar is the shortest in nearly every group.}
\end{figure}

\compactparagraph{Testbed.}
Our primary platform is a OnePlus 12 with a Snapdragon 8 Gen 3 SoC, an Adreno 750 GPU, and 16 GB of unified memory.
To assess portability (Section~\ref{sec:portability}), we also evaluate
a OnePlus 11 with a Snapdragon 8 Gen 2 SoC, an Adreno 740 GPU, and 16 GB of memory to test same-vendor generalization,
and a Google Pixel 8 with a Tensor G3 SoC, a Mali-G715 GPU, and 8 GB of memory to validate cross-vendor portability under tighter resource constraints.

\subsection{End-to-End Performance Comparison}
This section compares the end-to-end performance of \projectname with the state-of-the-art baselines on mobile GPUs 
across various context lengths and model precisions.

\compactparagraph{Average latency with fixed context length.}
We report average per-token {\it decoding} latency across four models and four baselines
under three Prompt--Output ($P$--$O$) windows: 768--512, 512--512, and 256--512,
where, e.g., 768--512 denotes a 768-token prompt followed by 512 autoregressively decoded tokens.
As shown in Figure~\ref{fig:eval-e2e-seq-combined},
\projectname achieves lower per-token latency than all baselines across all configurations.
Against {LLAMA.CPP}, the average speedups are
1.29$\times$ (FP16), 2.47$\times$ (W8A8), and 1.59$\times$ (W4A8);
against MNN-LLM, 1.55$\times$, 2.06$\times$, and 1.60$\times$;
against LiteRT-LM, 1.37$\times$, 1.84$\times$, and 1.63$\times$;
against MLC-LLM, 1.38$\times$ (FP16) and 1.56$\times$ (W4A8) (this framework does not support W8A8).
FP16 also benefits from our adaptive layouts, though the gains are more modest 
because conversion overhead is not a major factor here. 

\begin{figure}[t!]
  \centering
  \includegraphics[width=0.91\columnwidth]{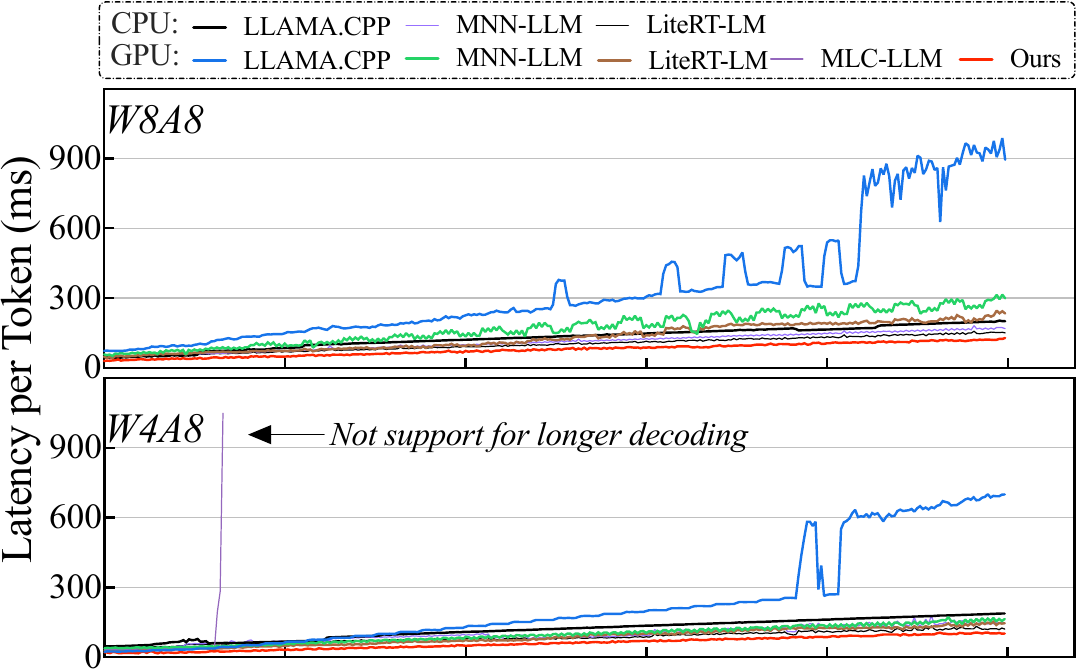}
  \caption{\textbf{Per-token decoding latency as sequence length grows (up to $10K$ tokens).} 
          MLC-LLM does not support W8A8 or long contexts at W4A8.
          Our method (red) achieves the lowest latency and reduces the stair-step spikes observed in baselines.}
  \label{fig:eval-e2e-token-combined}
  \vspace{-1em}
  \Description{Line charts of per-token latency versus sequence length at W8A8 and W4A8; GPU LLAMA.CPP shows large stair-step jumps while Ours stays lowest and nearly flat.}
\end{figure}

Notably, \projectname achieves 1.66$\times$ (W8A8) and 2.36$\times$ (W4A8) speedups over FP16,
recovering 83\% and 59\% of the theoretical 2$\times$/4$\times$ bounds.
The gap stems from non-quantized components (e.g., {\tt Softmax} and {\tt Normalization})
and fixed runtime overheads that are not reduced with bit-width.
In contrast, baselines capture far less: 
LLAMA.CPP achieves only 0.75$\times$ at W8A8 (even slower than FP16) and 1.56$\times$ at W4A8;
MNN-LLM reaches 1.25$\times$/1.86$\times$; 
LiteRT-LM reaches 1.14$\times$/1.79$\times$;
MLC-LLM attains 1.76$\times$ at W4A8.
This demonstrates that \projectname translates quantization potential into 
end-to-end speedups more effectively than prior systems. 

Moreover, \projectname's advantage widens in longer prompts:
against LLAMA.CPP, the average speedup grows from
1.62$\times$ (256--512) to 1.98$\times$ (768--512);
similar trends hold for MNN-LLM (1.61$\times$$\to$1.85$\times$),
LiteRT-LM (1.50$\times$$\to$1.68$\times$),
and MLC-LLM (1.34$\times$$\to$1.62$\times$).
Longer prompts intensify prefill computations and expand the KV cache,
increasing sensitivity to memory layout.
Baselines apply a fixed layout regardless of shape, 
incurring growing penalties,
whereas our cache-aware runtime adapts layouts as tensor shapes evolve
and our {conversion-hiding} execution mitigates the overhead.

{%
\compactparagraph{Working with a fixed latency budget.}
Under a fixed latency budget, \projectname can use its efficiency gains
to run higher-precision configurations and improve model quality.
Across the four models and three $P$--$O$ windows in
Figure~\ref{fig:eval-e2e-seq-combined}, \projectname at FP16 achieves
geometric-mean speedups of 1.75$\times$, 1.46$\times$, and 1.24$\times$
over LLAMA.CPP, MNN-LLM, and LiteRT-LM at W8A8, respectively.
MLC-LLM is omitted because it does not support W8A8.
Similarly, \projectname at W8A8 outperforms LLAMA.CPP, MNN-LLM,
LiteRT-LM, and MLC-LLM at W4A8 by 1.08$\times$, 1.07$\times$,
1.04$\times$, and 1.02$\times$, respectively.
Table~\ref{tab:eval-model-accuracy} confirms the corresponding quality
benefit: FP16 reduces perplexity by 0.13--2.24 points over W8A8,
while W8A8 reduces it by 0.31--1.79 points over W4A8.
Thus, \projectname can either reduce latency at fixed precision or
improve model quality under a fixed latency budget.
}

\compactparagraph{Latency scaling with increasing context length.}
Next, we analyze how speedups evolve with decoding length.
We fix $P$ and vary $O$ from 1 to $10K$.
Due to space constraints, we report results for a representative model (Llama-3.2-1B),
as other models exhibit similar trends.
We present results for the low-bit precisions W8A8 and W4A8.

Figure~\ref{fig:eval-e2e-token-combined} shows that
\projectname consistently outperforms all baselines as $O$ increases.
Our per-token latency remains flat, whereas all baselines exhibit periodic spikes
from increasingly inefficient layouts as the KV cache grows.
\projectname achieves average speedups of
3.25$\times$, 1.68$\times$, and 1.50$\times$, respectively,
over LLAMA.CPP, MNN-LLM, and LiteRT-LM at W4A8.
MLC-LLM degrades catastrophically at longer contexts (speedup over 43.81$\times$)
due to its static layout design;
we therefore exclude it from the primary long-context comparison.
At W8A8, the speedups are 3.96$\times$, 2.14$\times$, and 1.81$\times$ against LLAMA.CPP, MNN-LLM, and LiteRT-LM.
Notably, for these baselines the CPU backend often outperforms their own GPU backend
at low-bit precisions, owing to richer native support for low-precision integer arithmetic on CPUs.
Even against their best results across platforms, 
which happen to be the CPU backends,
\projectname on GPU still achieves substantial speedups:
2.11$\times$, 1.65$\times$, and 1.56$\times$ over LLAMA.CPP, MNN-LLM, and LiteRT-LM at W8A8,
and 1.74$\times$, 1.42$\times$, and 1.38$\times$ at W4A8. 

\begin{figure}[t!]
  \centering

  \includegraphics[width=0.91\columnwidth]{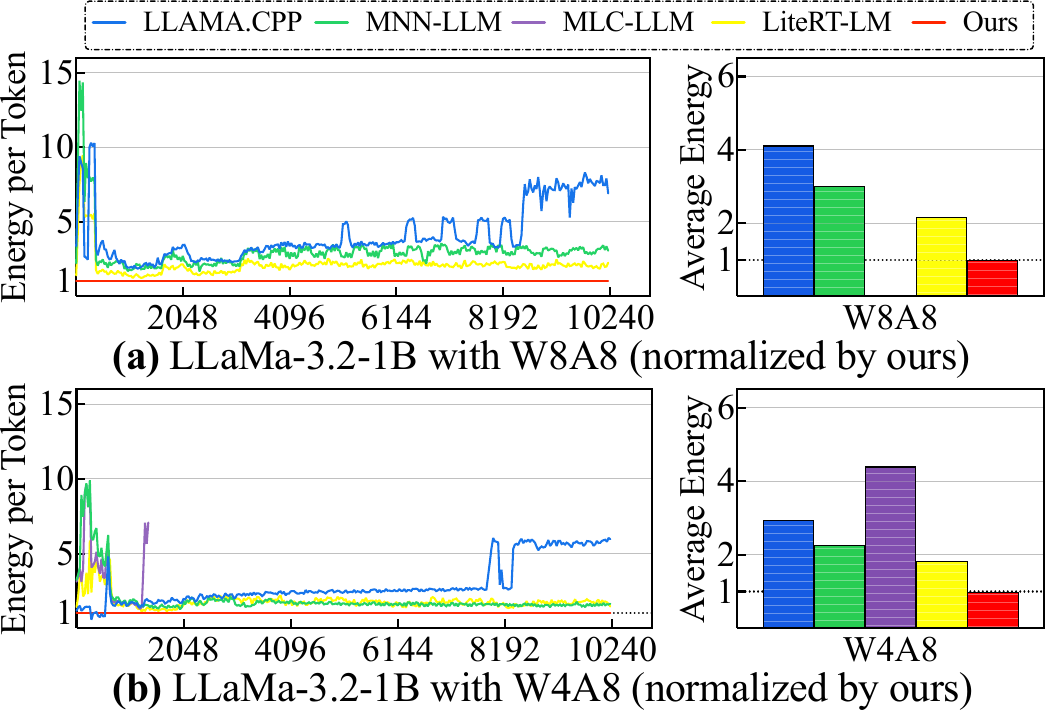}
  \caption{{\textbf{Normalized (by ours) per-token and average energy consumption for low-bit precisions.}}}
  \label{fig:eval-e2e-energy-combined}
  \Description{Energy per token versus sequence length and average energy bars for LLaMA-3.2-1B at W8A8 and W4A8, normalized to Ours; baselines use roughly 2 to 4 times more energy on average.}
\end{figure}

Speedups grow monotonically with $O$ as KV cache expansion amplifies layout sensitivity.
From $O{=}1$ to $O{=}10K$, the gain over LLAMA.CPP rises from
2.0$\times$ to 8.35$\times$ (W8A8) and 1.1$\times$ to 6.93$\times$ (W4A8);
over MNN-LLM, from 1.49$\times$ to 2.66$\times$ (W8A8) and 1.38$\times$ to 2.48$\times$ (W4A8);
over LiteRT-LM, from 1.24$\times$ to 2.03$\times$ (W8A8) and 1.21$\times$ to 1.98$\times$ (W4A8).
The trend is pronounced against MLC-LLM at W4A8,
where speedup escalates from 1.36$\times$ to 527$\times$,
as its static layout incurs increasingly severe penalties with KV cache growth.
These results confirm that our adaptive layout and integer-native execution
mitigate the overhead of growing context length.

\compactparagraph{Energy consumption per token and average profile.}
Figure~\ref{fig:eval-e2e-energy-combined} reports normalized per-token energy (left) 
and average energy (right) for Llama-3.2-1B under W8A8 and W4A8, 
with $O$ from 1 to $10K$ and a fixed prompt of 512.
\projectname reduces average per-token energy by 75.7\%/66.0\% (W8A8/W4A8) over LLAMA.CPP,
66.7\%/56.1\% over MNN-LLM, 
54.4\%/44.2\% over LiteRT-LM,
and 88.9\% over MLC-LLM at W4A8.
Consistent with the latency, our energy consumption remains stable as $O$ grows,
whereas baselines show rising energy overhead with longer sequences.

\subsection{Performance Breakdown and Analysis}\label{sec:evaluation-breakdown}

\begin{figure}[t!]
  \centering
  \includegraphics[width=0.96\columnwidth]{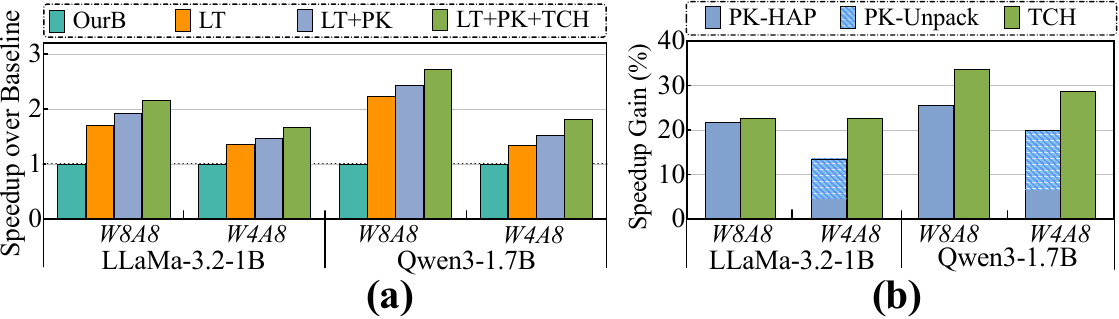}
  \vspace{-0.5em}

  \caption{\textbf{Optimization breakdown on mobile GPU.}
  (a) Cumulative speedup over {\tt OurB} (baseline = 1) as each
  optimization is enabled. {\tt LT}, {\tt PK}, and {\tt TCH} denote
  adaptive layout, quantization-aware packing, and type-conversion
  hiding, respectively. {(b) Fine-grained breakdown of
  the incremental speedup gains from {\tt PK} and {\tt TCH} over {\tt OurB}.
  The {\tt PK} gain is separated into hardware-aligned packing ({\tt PK-HAP}) and reduced INT4 unpacking ({\tt PK-Unpack}).
  }}
  
  \label{fig:eval-opt-breakdown}
  \Description{Bar charts of cumulative speedup as layout, packing, and conversion-hiding optimizations are enabled, and of the percentage gain contributed by packing and conversion hiding.}
\end{figure}

\compactparagraph{Optimization breakdown.}
Figure~\ref{fig:eval-opt-breakdown}(a) reports the cumulative speedup of each optimization
for two representative models (Llama-3.2-1B and Qwen3-1.7B)
under a fixed $P$--$O$ window of 512--512 at W8A8 and W4A8 precisions.
Other models exhibit similar trends and are omitted for brevity.
All bars are normalized to {\tt OurB}, a variant of \projectname with all key optimizations disabled;
each subsequent bar adds one optimization cumulatively.
{
Accordingly, the differences between adjacent bars quantify the incremental contribution of each component at each precision: {\tt OurB}$\rightarrow${\tt LT} for adaptive layout, {\tt LT}$\rightarrow${\tt LT + PK} for quantization-aware packing, and {\tt LT + PK}$\rightarrow${\tt LT + PK + TCH} for the type-conversion-hiding pipeline.}
At W8A8, {\tt LT} (adaptive layout) delivers 1.71$\times$ and 2.23$\times$ speedups
on Llama-3.2-1B and Qwen3-1.7B, respectively.
Adding {\tt PK} (quantization-aware packing) yields an additional 1.22$\times$ and 1.26$\times$;
{\tt TCH} (type-conversion hiding) contributes a further 1.23$\times$ and 1.34$\times$.
At W4A8, the corresponding incremental gains are
1.37$\times$ and 1.34$\times$ ({\tt LT}),
1.1$\times$ and 1.2$\times$ ({\tt PK}),
1.2$\times$ and 1.29$\times$ ({\tt TCH}).
Among all components, {\tt LT} contributes the largest improvement
by preserving data locality through layout adaptation,
enabling memory bandwidth to be utilized more effectively
(further quantified in the following analysis).

{
\compactparagraph{Quantization-aware packing breakdown.}
Figure~\ref{fig:eval-opt-breakdown}(b) further shows the incremental
gains over {\tt OurB} that are the cumulative speedups reported in
Figure~\ref{fig:eval-opt-breakdown}(a).
It decomposes {\tt PK} gain into hardware-aligned packing ({\tt PK-HAP})
and reduced sub-byte unpacking ({\tt PK-Unpack}), and reports {\tt TCH}
separately as the contribution of the subsequent conversion-hiding kernel
pipeline.
{\tt PK-HAP} captures the benefit of matching the stored operand order to the
RGBA texture lanes and dot-product instructions, whereas {\tt PK-Unpack}
isolates the reduction in shift-and-mask operations for packed INT4 values.
At W8A8, {\tt PK-HAP} contributes 21.9\% and 25.7\% on Llama-3.2-1B
and Qwen3-1.7B, respectively, while {\tt PK-Unpack} has no contribution
because INT8 values require no sub-byte extraction.
At W4A8, {\tt PK-HAP} contributes 4.63\% and 6.71\%, and
{\tt PK-Unpack} contributes 8.95\% and 13.2\%, respectively.
The subsequent {\tt TCH} optimization contributes 22.6\% and 33.7\%
at W8A8 and 22.6\% and 28.7\% at W4A8.
}

\begin{figure}[t!]
  \centering
  \includegraphics[width=0.92\columnwidth]{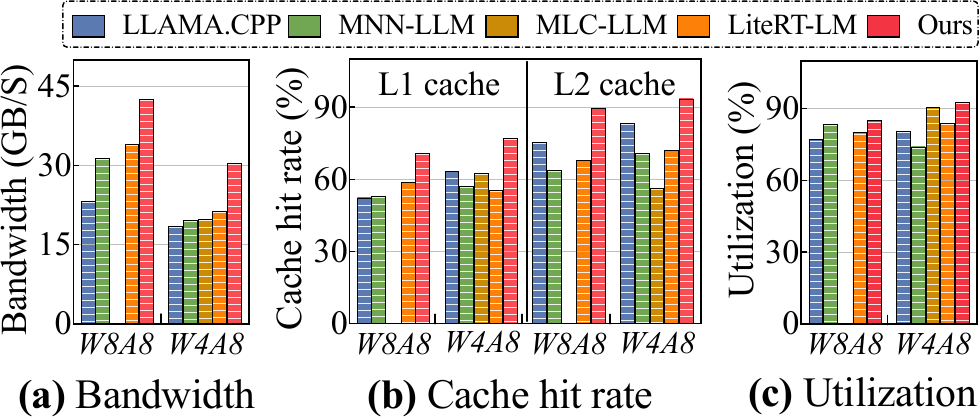}
  \caption{\textbf{Memory, cache, and ALU utilization analysis.}}
  \label{fig:eval-mem-cache-alu}
  \Description{Three bar charts comparing memory bandwidth, L1 and L2 cache hit rate, and ALU utilization across engines; Ours is highest in each metric.}
\end{figure}

{
The W8A8 packing gain therefore comes entirely from {\tt PK-HAP}.
Although the default W8A8 representation is byte-aligned,
its element order does not match the fixed lane order in which
a 32-bit dot-product instruction consumes the four values
returned by an RGBA texture fetch.
This mismatch requires lane rearrangement.
{\tt PK-HAP} instead stores each group in consumption order,
allowing the fetched values to feed the instruction directly.
For the profiled Llama-3.2-1B workload($P$--$O$=512--512, Adreno~750),
Figure~\ref{fig:eval-mem-cache-alu} shows that W8A8 exhibits
25.4\%--60.2\% higher measured bandwidth than W4A8
across baselines supporting both precisions,
with mean L1/L2 cache-hit-rate decreases of 3.9\%/6.3\%.
Given this higher memory pressure,
\projectname achieves a larger bandwidth gain at W8A8,
reaching up to 1.84$\times$ the measured baseline bandwidth,
compared with 1.65$\times$ at W4A8.
This contrast explains the larger {\tt PK-HAP} benefit in
Figure~\ref{fig:eval-opt-breakdown}(b),
while W4A8 gains more from reduced INT4 unpacking.
}

\compactparagraph{Memory, cache, and ALU analysis.}
The preceding breakdown quantifies each optimization's contribution;
we now profile the hardware-level metrics that explain these gains.
Figure~\ref{fig:eval-mem-cache-alu} reports memory bandwidth, cache behavior,
and ALU utilization measured via Snapdragon Profiler~\cite{snapdragon-profiler}.
Figure~\ref{fig:eval-mem-cache-alu}~(a) shows GPU memory bandwidth
at W8A8 and W4A8 (MLC-LLM is excluded at W8A8 as it does not support this precision).
\projectname achieves the highest bandwidth across both precisions,
reflecting the aligned access patterns produced by adaptive layouts
and operator-aware folding (Section~\ref{sec:design:layout}).
Figure~\ref{fig:eval-mem-cache-alu}~(b) shows cache hit rates:
\projectname achieves 13.6\%--18.6\% higher L1 and 10.4\%--37.3\% higher L2 hit rates than all baselines.

W4A8 yields higher hit rates than W8A8 due to the reduced memory footprint.
{
This also explains why W4A8 baselines can outperform W8A8 on mobile GPUs.
In the profiled configuration, relative to W8A8,
W4A8 reduces memory traffic by 39.8\%,
increases the L1 and L2 cache hit rates by 21.14\% and 12.3\%,
respectively, and increases unpacking overhead by 9.04\%.
For memory- and cache-sensitive decoding,
reduced data movement and improved cache locality
can outweigh the additional INT4 unpacking cost.
Conversely, the higher memory pressure of W8A8
increases the benefit of hardware-aligned packing,
consistent with the larger {\tt PK-HAP} contribution in Figure~\ref{fig:eval-opt-breakdown}(b).
}
The improved cache locality follows from stride-1 access preservation
through non-reduction-axis folding and length-aware layout selection.
Figure~\ref{fig:eval-mem-cache-alu}~(c) reports shader ALU utilization.
\projectname achieves the highest utilization at both precisions,
with improvements of up to 18.6\% at W8A8 and 7.9\% at W4A8.
This gain results from quantization-aware packing (reducing unpacking instructions
by $\approx$80\%) and conversion-hiding execution, which reduce unpacking work
and remove type conversion from the compute critical path, respectively.

\begin{figure}[t]
  \centering
  \begin{minipage}[t]{0.34\columnwidth}
    \vspace{0pt}
    \centering
    \includegraphics[width=\linewidth]
    {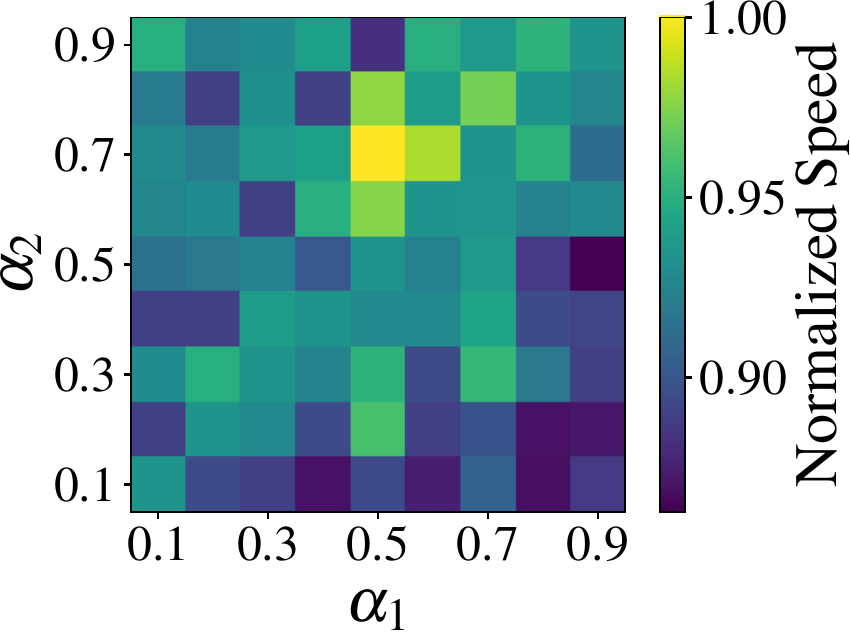}
  \end{minipage}\hfill
  \begin{minipage}[t]{0.63\columnwidth}
    \vspace{0pt}
    \centering
    {%
      \scriptsize
      \fontfamily{ptm}\selectfont
      \setlength{\tabcolsep}{1.3pt}
      \renewcommand{\arraystretch}{1.40}
      \resizebox{\linewidth}{!}{%
        \begin{tabular}{lrrrr}
          \toprule
          GPU &
          \makecell{Oracle \\search time} &
          \makecell{LEP \\search time} &
          \makecell{Mean\\slowdown $\downarrow$} &
          \makecell{P95\\slowdown $\downarrow$} \\
          \midrule
          Adreno-750 & $1,414$s & $45.5$ms & $1.86$\% & $7.27$\% \\
          Adreno-740 & 1,578s & $52.9$ms & 2.36\% & 5.83\% \\
          Mali-G715  & 2,747s & 95.2ms & 3.89\% & 10.40\% \\
          \bottomrule
        \end{tabular}%
      }
    }
  \end{minipage}

  \makebox[0.34\columnwidth][c]{{\small\bfseries (a)}}\hfill
  \makebox[0.63\columnwidth][c]{{\small\bfseries (b)}}
  \Description{A heatmap of normalized speed over LEP weights alpha 1 and alpha 2, peaking near 0.5 and 0.7, and a table of oracle versus LEP search time and slowdown on three mobile GPUs.}
  \caption{\textbf{LEP calibration and selection analysis.}
  (a) Impact of different LEP weights $(\alpha_1,\alpha_2)$.
  {(b) LEP search overhead and performance gap to oracle across mobile GPUs.}}
  \label{fig:lep-analysis}
\end{figure}

\compactparagraph{{Analysis of LEP weights.}}\label{sec:evaluation-lep}
Figure~\ref{fig:lep-analysis}(a) presents the sensitivity of LEP
to different $\alpha_1$ and $\alpha_2$ settings on our primary testbed. 
The specific values are obtained from the offline calibration 
(in Section~\ref{sec:design:formulation}), which completes in less than 10 minutes.
The parameters are model-independent, and
the best-performing setting ($\alpha_1 = 0.5$, $\alpha_2 = 0.7$) is selected and used across all models.
Even drastically different parameter values result in only about a 10\%--15\% performance degradation.
Incorporating automated tuning strategies (e.g., Bayesian optimization) 
to further streamline the calibration process is left as future work.
Figure~\ref{fig:lep-analysis}(b) compares LEP with exhaustive
oracle search across representative mobile GPUs. Oracle search benchmarks every
candidate layout, requiring 1413.6--2747.3~s across the three GPUs. LEP completes
layout selection in only 45.5--95.2~ms, reducing search time by more than four
orders of magnitude. The selected layouts remain close to the oracle, with mean
slowdowns of 1.86--3.89\% and P95 slowdowns of 5.83--10.40\%. 
These results demonstrate stable selection quality across mobile-GPU architectures.

\subsection{Portability Evaluation}
\label{sec:portability}

\begin{figure}[t!]
  \centering
  \includegraphics[width=0.95\columnwidth]{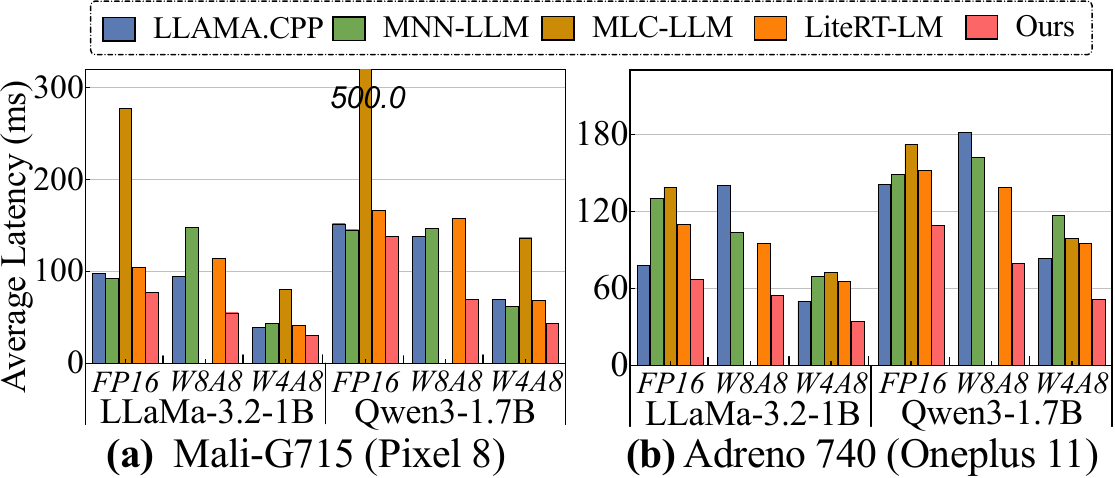}
  \caption{{\bf Portability evaluation.} Empty bars indicate that the framework does not support the configuration due to missing operator support or insufficient device memory.}
  \label{fig:eval-portability-e2e-s512}
  \Description{Bar charts of average latency on a Mali-G715 and an Adreno 740 GPU for two models at three precisions; Ours is the lowest bar in each group.}
\end{figure}

We assess portability on two additional platforms:
the OnePlus 11 (same vendor, earlier generation) and the Google Pixel 8 (cross-vendor, more constrained resources).
Figure~\ref{fig:eval-portability-e2e-s512} reports per-token latency
for Llama-3.2-1B and Qwen3-1.7B at FP16, W8A8, and W4A8 under $P$--$O$=512--512.
\projectname runs all model--precision combinations on both devices.
At low-bit precisions, \projectname achieves 1.3$\times$--2.9$\times$ speedups on OnePlus 11
and 1.4$\times$--3.6$\times$ on Pixel 8 (average 1.7$\times$ vs.\ 1.9$\times$,
as tighter resources amplify the benefit of our optimizations).
At FP16, speedups of 1.1$\times$--2.1$\times$ and 1.1$\times$--3.6$\times$ 
confirm that adaptive layouts alone provide significant benefits.

\section{Related Work}
\label{sec:related}

\compactparagraph{Quantization for LLMs.}
Classical quantization methods for deep learning have evolved into LLM-specific schemes
that preserve model accuracy at aggressively low-bit widths~\cite{shen2025quartdepth}.
LLM.int8~\cite{dettmers2022gpt3} halves memory relative to FP16 with an outlier-aware 8-bit scheme.
SmoothQuant~\cite{xiao2023smoothquant} migrates activation scaling into weights for W8A8 quantization.
GPTQ~\cite{frantar2022gptq} performs one-shot weight quantization for accurate 3--4 bit deployment.
MobileQuant~\cite{mobilequant2024} develops mobile-friendly quantization algorithms 
for on-device applications, which is complementary to our setting.
Quantization has also been applied to reduce KV cache costs:
KVQuant~\cite{hooper2024kvquant} provides calibration-based 4-bit KV quantization,
and Q-Hitter~\cite{zhang2024qhitter} incorporates token-oracle signals for better compression.
SNIP~\cite{pan2026snip} explores sub-byte mixed-precision for LLM training, 
while AWQ~\cite{lin2024awq} rescales salient channels for 4-bit quantization.
These works primarily target server-class platforms and their specialized low-bit instruction sets, 
such as CUDA INT4/INT8 and AVX-512 VNNI.
Mainstream mobile GPUs (e.g., Adreno, Mali) lack such dedicated instruction support;
our work is designed for these architectures.

\compactparagraph{Low-bit computation optimizations.}
A line of work on precision-scalable hardware~\cite{sharma2018bit,judd2016stripes,delmas2017dynamic,sharify2018loom,albericio2017bit}
scales execution width to operand precision, but largely targets custom accelerators.
T-MAC~\cite{t-mac} and BitNet.cpp~\cite{bitnet} explore table-lookup-based low-bit computation on edge-device CPUs,
but their access patterns map poorly to GPUs due to uncoalesced memory accesses
and misalignment with GPU memory hierarchies.
PMPD~\cite{pmpd2025} focuses on mixed-precision scheduling, 
adapting precision across inference phases and over the generated sequence 
to improve efficiency on resource-constrained hardware.
On the server side, Atom~\cite{zhao2024atom} and QServer~\cite{qserver} 
integrate mixed-precision quantization with CUDA tensor-core kernels unavailable on mobile GPUs.
At format and language level,
M2XFP~\cite{hu2026m2xfp} proposes a metadata-augmented microscaling format for low-bit quantization,
and Tilus~\cite{ding2026tilus} provides a tile-level GPU programming language for low-precision computation.
Other closely related work generates quantized kernels whose {\em lane-interleaving}
or {\em bit-packing layouts}
are dictated by specific SIMD instructions
(e.g., ARM NEON~\cite{cowan2020automatic}, x86 VNNI~\cite{shen2025efficient}, and Tensor Core MMA~\cite{frantar2025marlin}),
and are resolved {\em statically} at compile time or during offline preprocessing.

\compactparagraph{LLM serving and compiler frameworks.}
Several frameworks target LLM inference on server GPUs~\cite{tensorrtllm,fastertransformer,xiao2023efficient,pope2023efficiently,zhu2026easerve}.
DeepSpeed-Inference~\cite{aminabadi2022deepspeed} combines kernel fusion, tensor parallelism, and quantization-aware execution.
FlashAttention~\cite{dao2022flashattention,dao2023flashattention} improves attention kernel utilization through optimized work partitioning.
SGLang~\cite{zheng2024sglang} and Vidur~\cite{agrawal2024vidur} explore scheduling and serving strategies.
oFFN~\cite{song2026offn} accelerates inference with an outlier-aware structured FFN design.
For KV cache management,
vLLM~\cite{kwon2023efficient} reduces fragmentation via paged attention.
Paged KV-cache management and speculative decoding have been developed
primarily for server-class LLM serving,
where memory management in high-throughput request processing are central.
In contrast, \projectname targets efficient low-bit decoding
on resource-constrained mobile GPUs,
but these techniques remain complementary.
A paged KV cache could be integrated by treating pages or page groups
as layout-managed segments.
Speculative decoding would require additional KV-state management;
we leave their implementation and evaluation to future work.
Data layout also matters for efficient DNN execution~\cite{niu2024smartmem}.
TVM~\cite{chen2018tvm} optimizes graph-level fusion and layout selection;
Ansor~\cite{zheng2020ansor} and TMModel~\cite{guan2025tmmodel} search over loop transformations and tiling with learned cost models.
LADDER~\cite{ladder} transforms computations into statically optimized low-precision pipelines
with memory-layer-specific tensor transformations.
Triton~\cite{tillet2019triton} enables tile-centric code generation with near-vendor-library performance.

{These compiler systems primarily optimize code generation and
static schedules for predetermined layouts. \projectname instead uses LEP to
select quantization-aware layouts dynamically across decoding phases and couples the selected
layouts with a type-conversion-hiding execution pipeline.}

\compactparagraph{On-device LLM inference.}
While earlier mobile inference systems focused on convolution-heavy workloads~\cite{lee2019mobisr,yeo2020nemo,zhang2021nn,jeong2022band,niu2020patdnn},
the shift to LLMs has driven a new generation of on-device runtimes~\cite{ollama,llamafile,wang2024whisperflow,alizadeh2024llm}.
LLAMA.CPP~\cite{llamacpp} and LiteRT-LM~\cite{litert-lm} provide cross-platform runtimes 
leveraging vector instructions and low-bit formats.
MLC-LLM~\cite{mlc-llm}, built on TVM, enables low-bit LLM execution on mobile GPUs.
However, as our evaluation reveals, these frameworks inherit static layout assumptions
that do not adapt to the dynamic memory access patterns of autoregressive decoding,
leaving a performance gap on mobile GPUs.
SoD$^2$~\cite{niu2024sod2} and Transformer-Lite~\cite{transformerlite2024} target 
dynamic-shape inference and memory planning for mobile deployment, 
achieving significant speedups through static shape propagation and operator-level optimizations.
FlashMem~\cite{shu2026flashmem} optimizes the mobile GPU memory hierarchy for LLM workloads,
though it targets floating-point operations and does not handle the growing KV cache during decoding.
Neuralink~\cite{wang2025neuralink} reduces on-device LLM compute through neuron co-activation linking,
which is orthogonal to the layout and type conversion overhead we focus on.
Production frameworks including
ExecuTorch~\cite{executorch} and ONNXRuntime~\cite{onnixruntime}
primarily target mobile CPUs, with limited low-bit GPU backend support.
Other on-device efforts target NPUs and heterogeneous SoC execution,
including PowerInfer~\cite{song2024powerinfer,xue2024powerinfer}, Ripple~\cite{wang2024ripple},
LLM.NPU~\cite{xu2025fast}, HeteroLLM~\cite{chen2025heterollm}, and T-MAN~\cite{t-man},
but do not address the GPU-oriented, low-bit decoding challenges we focus on;
extending our optimizations to such heterogeneous scenarios is a promising future direction.
{
Overall, \projectname's distinction lies not in any individual
layout, packing, or pipelining technique, but in their sequence-aware
coordination for mobile-GPU decoding. 
The underlying framework, LEP,  dynamically makes layout and packing decisions as the KV cache grows, 
while the conversion-hiding pipeline preserves these gains during execution.}

\section{Discussion and Future Work}

{
\compactparagraph{Generality across execution environments.}
At the LEP level, supporting a new mobile GPU requires updating the
supported layout candidates and recalibrating the device-specific weights
once offline, as detailed in Section~\ref{sec:design:formulation} and
evaluated across GPUs in Section~\ref{sec:evaluation-breakdown}.
}
{For \textit{a heterogeneous SoC} (i.e., with CPU, GPU, and NPU), LEP can be extended by
scoring backend--layout pairs and incorporating an inter-backend transition
term:}
{
\begin{equation*}
\begin{aligned}
\mathrm{LEP}_{b'\rightarrow b}(\mathcal{L},t)
={}&\alpha_{1,b}\frac{\mathrm{Bytes}_{b}(\pi,\rho,\phi,t)}{\mathrm{BW}_{b}}\\
&+\alpha_{2,b}\,\mathrm{UnpackOps}_{b}(\phi)
+T_{\mathrm{trans}}(b'\!\rightarrow\! b,\mathcal{L},t).
\end{aligned}
\end{equation*}
}

{Here, $b',b\in\{\mathrm{CPU},\mathrm{GPU},\mathrm{NPU}\}$ denote
the source and target backends. The current GPU implementation assumes 2D
texture and RGBA access, GPU-oriented packing, and low-bit dot-product
instructions. Extending LEP to CPUs and NPUs would preserve the cost-model
structure but require backend-specific layouts, memory models, and execution
primitives (i.e., instruction sets). 
The transition term $T_{\mathrm{trans}}$ captures synchronization,
data movement, and other required layout conversion between backends; 
it is zero when $b'=b$.
The heterogeneous extension would retain LEP's lightweight selection procedure.
For a fixed source backend $b'$, LEP evaluates
$\sum_b|\mathcal{C}_b|$ backend--layout candidates.
Because the backend-specific LEP parameters are precomputed, each candidate is evaluated with lightweight computation.}

{\textit{Concurrent workloads} can change the effective backend and bandwidths,
causing parameters obtained through offline calibration to become suboptimal. 
A {\em contention-aware}  extension could update these
parameters at decoding-phase boundaries based on resource-availability
information provided by the runtime scheduler and then rescore the affected
candidates. Extending to contention-aware heterogeneous execution is a promising direction.}

\compactparagraph{Extending to new quantization schemes.}
\projectname currently evaluates W8A8 and W4A8 using a group size of 32; 
however, its layout-selection and packing formulation is not specific to these configurations.
Supporting an additional quantization format requires incorporating compatible packing candidates 
and extending the LEP cost components to capture format-specific overheads.
Specifically, the $\mathrm{Bytes}$ term could account for quantized data, 
whereas the $\mathrm{UnpackOps}$ term accounts for bit extraction, operand rearrangement, and type conversion.
These costs may depend on both group size and bit width: 
group size determines dequantization granularity, 
while bit width determines packing density, unpacking cost, and the availability of compatible integer primitives.
For each target device, a packing candidate is associated with a supported execution path; 
when the packed representation cannot be consumed directly by a native primitive, 
LEP can further incorporate the required unpacking and conversion costs.
LEP then ranks the extended set of candidates using the resulting cost model.
We leave additional quantization formats to future work.

\section{Conclusion}
We presented \projectname, 
which addresses the unrealized potential of low-bit LLM inference on mobile GPUs 
through adaptive KV cache layouts, hardware-aligned packing, and an integer-native pipeline 
that defers type conversion outside the compute hot path.
Extensive evaluation demonstrates that \projectname delivers significant speedups
(e.g., 1.78$\times$ over LLAMA.CPP on average with fixed context length)
and substantial per-token energy reductions (up to 88.9\%),
while effectively eliminating latency spikes during KV cache growth
to ensure predictable long-context decoding.
Future work includes combining other low-bit quantization schemes with pruning co-optimization 
and support for concurrent workloads on heterogeneous mobile hardware (NPUs, TPUs).

\begin{acks}
The authors would like to thank the anonymous reviewers and the shepherd for their valuable and constructive feedback.
This work was supported in part by the National Science Foundation (NSF) under the awards of 
CCF-2428108, CCF-2333895, OAC-2403090, CSR-2341378, IIS-2540429, CNS-2555319, and CCF-2553684.
Any errors and opinions are not those of the NSF and are attributable solely to the author(s). 
\end{acks}

\bibliographystyle{ACM-Reference-Format}
\balance
\bibliography{reference}

\end{document}